\documentclass[letterpaper]{JHEP3}

\usepackage{bm,amsmath,amssymb,mathrsfs,slashed}
\usepackage{epsfig,graphics,graphicx,subfigure}
\usepackage{cite}
\usepackage[latin1]{inputenc}

\long\def\comment#1{ }
\newcommand{\eqnum}[1]{Eq.~\eqref{#1}}

\newcommand{\beq}{\begin{eqnarray}}
\newcommand{\eeq}{\end{eqnarray}}
\newcommand{\nn}{\nonumber\\}
\newcommand{\dif}{{\rm d}}
\newcommand{\rmd}{{\rm d}}
\newcommand{\rme}{{\rm e}}
\newcommand{\rmi}{{i}}

\newcommand{\rmJ}{{\rm J}}
\newcommand{\rmH}{{\rm H}}
\newcommand{\rmK}{{\rm K}}
\newcommand{\rmI}{{\rm I}}

\newcommand{\del}{\partial}

\newcommand{\lan}{\langle}
\newcommand{\ran}{\rangle}

\newcommand{\order}[1]{\mcal{O}{(#1)}}
\newcommand{\mcal}{\mathcal}

\newcommand{\bp}{b_{\perp}}

\newcommand{\hh}{\tilde{h}}

\usepackage{accents}
\makeatletter
\def\cc@accentPhantom#1#2{%
\begingroup
\cc@code=#1\relax
\let\math@bgroup\@empty
\def\math@egroup##1{\cc@setchar##1}%
\cc@palette\cc@@accentPhantom{#2}%
\cc@nuc
\endgroup%
}
\def\cc@@accentPhantom#1#2{%
\let\cc@style=#1%
\cc@fetch{#2}%
\mathaccent\cc@code{%
\ifcc@more#2\else
\cc@phant{#2}%
\gdef\cc@nuc{}%
\fi
\kern\cc@skew%
}%
\kern-\cc@skew
\kern\cc@wd%
}
\gdef\dacute#1{%
\rlap{$\mkern-2.0mu\protect\cc@accentPhantom{"7013}{#1}$}%
\rlap{$\mkern2.0mu\protect\cc@accentPhantom{"7013}{#1}$}%
#1%
}
\makeatother

\title{Shockwaves and deep inelastic scattering within the gauge/gravity
duality}

\author{E.~Avsar,$^a$ E.~Iancu,$^a$ L.~McLerran$^b$ and D.N.~Triantafyllopoulos$^c$\\
\!\!$^a$Institut de Physique Th\'{e}orique de Saclay, F-91191 Gif-sur-Yvette, France \\
\!\!$^b$ RIKEN BNL Research Center and Physics Department,
Brookhaven National Laboratory, Upton, NY 11973, USA\\
\!\!$^c$Institut f\"{u}r Theoretische Physik, Technische Universit\"{a}t Wien,
1040 Vienna, Austria\\
E-mail: \email{Emil.Avsar@cea.fr, Edmond.Iancu@cea.fr,
mclerran@bnl.gov, dionysis@hep.itp.tuwien.ac.at}}

\abstract{Within the gauge/gravity correspondence, we discuss the general
formulation of the shockwave metric which is dual to a `nucleus' described
by the strongly--coupled ${\mathcal N}=4$ SYM theory in
the limit where the number of colors $N_c$ is arbitrarily large. We emphasize
that the `nucleus' must possess $N_c^2$ degrees of freedom per unit volume, so
like a finite--temperature plasma, in order for a supergravity description
to exist. We critically reassess previous proposals for introducing transverse
inhomogeneity in the shockwave and formulate a new proposal in that sense,
which involves no external source but requires the introduction of
an `infrared' cutoff which mimics confinement. This cutoff however
plays no role when the shockwave is probed by a highly virtual projectile,
so like in deep inelastic scattering. We consider two such projectiles,
the dilaton and the $\mcal{R}$--current, and compute the respective
structure functions including unitarity corrections. We find that there are no
leading--twist contributions to the structure functions at high virtuality,
meaning that there are no point--like constituents in the strongly coupled
`nucleus'. In the black--disk regime at low virtuality, the structure
functions are suggestive of parton saturation with occupation numbers of
order one. The saturation momentum $Q_s$ grows with the energy like $Q_s^2
\sim 1/x$ (with $x$ the Bjorken variable), which is the hallmark of graviton
exchanges and is also necessary for the fulfillment of the energy--momentum
sum rules.
}

\begin{document}
\setcounter{page}{1}

\section{Introduction}
\label{sect-Int} \setcounter{equation}{0}

Some of the experimental discoveries at RHIC, notably the unexpectedly
large medium effects known as elliptic flow and jet quenching, led to the
suggestion that the deconfined hadronic matter produced in the
intermediate stages of a heavy ion collision might be strongly
interacting \cite{Gyulassy:2004zy,Muller:2007rs}. This observation
triggered a large theoretical activity aiming at understanding the
properties of a strongly--coupled plasma via string theory methods,
within the AdS/CFT correspondence (see, e.g., the review papers
\cite{Son:2007vk,Iancu:2008sp,Gubser:2009sn} and refs. therein). Part of
this activity concentrated on the `dual' formulation of a heavy ion
collision at strong coupling. The main motivation for that problem is
less in the description of the nuclear wavefunctions by themselves --- a
real ultrarelativistic nucleus in QCD is rather described via weak
coupling methods, because of the high parton density produced by the
high--energy evolution \cite{CGCreviews} ---, but rather in understanding
the late stages of the collision and, especially, study thermalization.
If the deconfined matter produced after a heavy ion collision is
effectively strongly coupled, then one may hope that the subsequent
dynamics leading to thermalization is not very sensitive to the
perturbative early stages, and thus it can be studied by simulating a
full collision in AdS/CFT. Such a study may help explaining one of the
main puzzles left by RHIC, which is the very short thermalization time
suggested by the elliptic flow data.

The AdS/CFT correspondence
\cite{Maldacena:1997re,Gubser:1998bc,Witten:1998qj} deals with the
conformally symmetric ${\mathcal N}=4$ supersymmetric Yang--Mills (SYM)
theory, and not directly with QCD. But thus is probably not a major
impediment so long as we are interested in the temperature range
$2T_c\lesssim T \lesssim 5T_c$ (the relevant range for heavy ion
experiments at RHIC and LHC), where the QCD plasma itself appears to be
nearly conformal, as shown by lattice studies \cite{Cheng:2007jq}. For
practical reasons, all the studies so far were limited to the
`supergravity approximation' valid when the number of colors $N_c$, and
hence also the `t Hooft coupling $\lambda=g^2N_c$, are arbitrarily large.
(The gauge coupling $g$ is assumed to be small, $g\ll 1$, but fixed.) In
this approximation, a plasma or a nucleus are described as modifications
of the asymptotically AdS$_5$ geometry that the string theory lives in,
and the dynamics reduces to classical gravity in this curved space--time.
In order for the respective metric perturbation to survive as
$N_c\to\infty$, one needs to assume that the system under consideration
possesses $N_c^2$ degrees of freedom per unit volume. This is indeed the
case for the ${\mathcal N}=4$ SYM plasma at finite temperature, whose
dual description is a black--brane in AdS$_5$ \cite{Witten:1998zw}, but
it would not be true for a QCD--like nucleus, which is built with
colorless `hadrons'.

Still, if we are merely interested in studying thermalization, then one
can replace the colliding nuclei with finite--size slices of plasma,
represented as slices of `black--holes' (more properly, black branes) in
AdS$_5$. In the center--of--mass frame, where these `nuclei' (prior to
the collision) are highly energetic and hence strongly Lorentz
contracted, the black--hole slices appear as shockwaves directing against
each other. The ultimate objective is to explicitly compute the
scattering between two such shockwaves by solving Einstein equations, and
thus investigate the evolution towards a black brane at late stages,
which is synonymous of thermalization. Several steps have been already
done in that sense, with interesting results, and the dynamical formation
of a black hole has been already seen within supergravity calculations
\cite{Janik:2005zt,Grumiller:2008va,Gubser:2008pc,AlvarezGaume:2008fx,Lin:2009pn,Gubser:2009sx,Figueras:2009iu,Chesler:2008hg,Chesler:2009cy,Beuf:2009cx}.

But before attacking this complicated problem, it is the very formulation
of the gravity dual of a nucleus as a shockwave which still meets with
difficulties in the literature. The would--be natural recipe for building
such a shockwave in view of the discussion above --- namely, start with a
slice of the plasma with finite longitudinal width $L\gg 1/T$ in the
plasma rest frame and then boost the dual `black--hole slice' metric up
to a large Lorentz factor $\gamma\gg 1$ --- is not fully satisfactory at
a conceptual level, since a `slice of a black--hole' in AdS$_5$ is not an
exact solution to Einstein equations. Yet, as we shall demonstrate in
Appendix \ref{app-bhsw}, this approximate procedure has the merit to
generate a shockwave metric which is similar to the one that we shall
propose via different considerations, in Sect.~2.

An important related problem is that of the conformal symmetry breaking:
this was already necessary in order to built hadronic--like bound states
in the ${\mathcal N}=4$ SYM theory
\cite{Polchinski:2000uf,Polchinski:2001tt,Erlich:2005qh,Karch:2006pv,Brodsky:2008pg},
and one can hardly see how one could construct a nuclear shockwave
without introducing any scale in the problem. Yet, the original proposal
in that sense, due to Janik and Peschanski \cite{Janik:2005zt}, seems not
to require any such a scale. As we shall argue in Sect.~2, such a scale
must nevertheless be added by hand to that construction, in the form of a
cutoff on the radial dimension\footnote{Throughout this paper, we will be
using Fefferman--Graham coordinates (see Sect.~2 for precise
definitions), with the radial distance denoted as $z$. Thus by `large
radial distances' we mean large separations from the Minkowski boundary
of AdS$_5$, which in these coordinates lies at $z=0$.} of AdS$_5$ ---
either a `hard wall', \emph{i.e.} a sharp radial cutoff at $z=1/\Lambda$
\cite{Polchinski:2001tt}, or a `soft wall', like in
Ref.~\cite{Karch:2006pv}. This cutoff corresponds to an infrared cutoff
$\sim\Lambda$ in the ${\mathcal N}=4$ SYM theory and mimics the
confinement physics of QCD. Incidentally, the metric produced by boosting
a slice of the plasma is identical to that by Janik and Peschanski near
the Minkowski boundary (for $z\ll 1/T$), but it naturally involves a
radial cutoff, namely the black--hole horizon at $z\sim 1/T$ (see
Appendix A).

The need for a radial cutoff in AdS$_5$ can be also seen via a different
argument, which is internal to the supergravity construction: without
such a cutoff, the shockwave metric develops a genuine space--time
singularity at $z\to\infty$, and not just a coordinate singularity. This
has been recognized in the literature \cite{Podolsky:1997ik} for the
analog of the Janik--Peschanski metric in 4 dimensions, known as the
Kaigorodov space--time \cite{Kaigorodov}, but a similar argument holds in
5 dimensions. The argument is quite subtle, since the singularity at
$z\to\infty$ does not show up in any of the curvature scalars computed
from this metric, but is of a more general type, known as
$g$--incompleteness \cite{Hawking}.

The issue of the radial cutoff becomes even sharper when trying to
introduce {\em transverse inhomogeneity} in the shockwave, corresponding
to a profile for the nuclear matter distribution in the plane transverse
to the collision axis. The two types of proposals in that sense that we
are aware of
\cite{Gubser:2008pc,AlvarezGaume:2008fx,Lin:2009pn,Gubser:2009sx,Beuf:2009mk}
use external sources to break down conformal symmetry. Besides the
difficulty to motivate the $N_c^2$--scaling for the strength of these
sources, these proposals have some unwanted features, which make them
unrealistic from the viewpoint of QCD. The proposal in
Ref.~\cite{Gubser:2008pc} uses a pointlike source located in the bulk of
AdS$_5$ to produce a `nuclear' energy--momentum distribution which falls
off at large transverse distances according to a power law, instead of
the exponential law expected in the presence of confinement. Accordingly,
when used for scattering problems, this shockwave yields cross--sections
which violate the Froissart bound. The proposal in
Ref.~\cite{Beuf:2009mk} is even more objectionable, in that one is
mistreating the `ultraviolet' (high--momentum) part of the `nuclear'
wavefunction. This proposal involves a source on the boundary of AdS$_5$,
hence the corresponding metric perturbation does not vanish when
approaching the boundary. Via the UV/IR correspondence
\cite{Susskind:1998dq,Peet:1998wn,Polchinski:2002jw,Brodsky:2008pg,HIM2,HIM3},
this means that the spectrum of the quantum modes included in the dual
`nuclear' wavefunction is flat in the high--momentum limit, rather than
rapidly falling down, as it should on physical grounds.

On the other hand, the analysis in Ref.~\cite{Beuf:2009mk} has the virtue
to have identified a class of exact solutions to Einstein equations of
the shockwave type, which allow for a generic inhomogeneity in the
transverse plane. This analysis lies at the basis of our shockwave
proposal in this paper. One can succinctly describe our proposal as
follows: among the two general solutions found in \cite{Beuf:2009mk} and
which involves the modified Bessel functions $\rmK_2$ and, respectively,
$\rmI_2$, we shall discard the solution $\propto \rmK_2$ that was adopted
in Ref.~\cite{Beuf:2009mk} and keep the other solution $\propto \rmI_2$.
The latter is the generalization of the shockwave metric by Janik and
Peschanski \cite{Janik:2005zt} (which is homogeneous in the transverse
plane) to a generic transverse energy--momentum distribution. This
solution vanishes near the boundary ($z\to 0$) like $z^4$, meaning that
the spectrum of the modes included in the dual `nucleus' has an
acceptable high--momentum tail $\propto 1/Q^4$. On the other hand, this
solution would exponentially blow up at large values of $z$, but this is
not a problem since, as previously argued, the radial dimension must be
anyway supplemented with a cutoff at $z= 1/\Lambda$. From the perspective
of the boundary gauge theory, this scale $\Lambda$ plays several roles,
so like in QCD: it acts as an infrared cutoff for quantum fluctuations,
it sets the scale for energy--momentum density in the `nucleus', and also
the characteristic scale for transverse inhomogeneity.

The introduction of this radial cutoff is clearly an {\em ad hoc}
procedure (it spoils the exactness of the solution) and any quantity
which is sensitive to the details of this procedure is model--dependent.
This is probably the case for the collision between two such shockwaves,
but this is hardly a surprise: already in QCD, total hadronic
cross--sections are dominated by soft interactions and thus are sensitive
to the physics of confinement. But even in that case, one may hope that
the thermalization process in the late stages of the collision is less
sensitive to the details of the cutoff. On the other hand, the collision
between this `nucleus' and a `hard' probe with high transverse resolution
$Q\gg \Lambda$ is probing the shockwave metric only at small radial
distances $z\lesssim 1/Q\ll 1/\Lambda$, and hence it is completely
insensitive to the infrared cutoff. Once again, this is similar to QCD
where deep inelastic scattering at high--$Q^2$ is insensitive to
confinement.

The above considerations are illustrated by our calculations of deep
inelastic scattering (DIS) off the shockwave, which represent most of the
material in this paper. The first calculations of DIS at strong coupling
within the gauge/string duality referred to some other types of targets:
a `glueball' bound state
\cite{Polchinski:2002jw,HIM1,BallonBayona:2007qr,BallonBayona:2007rs}
(whose gravity dual is a normalizable dilaton state in AdS$_5$) and a
finite temperature plasma (as represented by the AdS$_5$ black--brane)
\cite{HIM2,HIM3}. More recently,
Refs.~\cite{Albacete:2008ze,Mueller:2008bt} presented the corresponding
calculations for a shockwave target (a `nucleus', or a `plasma slice'),
but their respective results appear to disagree with each other. The main
difference refers to the energy dependence of the saturation momentum
$Q_s$, which is the characteristic scale for the onset of unitarity
corrections: for $Q\gg Q_s$, the scattering is weak, whereas for
$Q\lesssim Q_s$ it reaches the unitarity bound, or `black disk', limit.

Already at weak coupling, $Q_s$ is known to grow quite fast with the
energy, as an inverse power of the Bjorken $x$ variable, due to the rapid
increase in the gluon density at small $x$ via bremsstrahlung
\cite{CGCreviews}. At strong coupling, one expects parton branching to be
faster and to `quasi--democratically' divide the energy among the
daughter partons
\cite{Polchinski:2002jw,HIM1,HIM2,Lin:2007fa,Strassler:2008bv,Hofman:2008ar}.
Accordingly, all partons should fragment down to very small values of $x$
and then the saturation momentum should grow  with $1/x$ even faster.
This is indeed what one found in Ref.~\cite{HIM1,HIM2,Mueller:2008bt} for
various types of hadronic targets: dilaton, infinite plasma, and a plasma
slice. Moreover, the energy dependence emerging from these calculations
looks very natural from the viewpoint of supergravity: the scattering
proceeds via multiple graviton exchanges, so the saturation momentum
inherits the energy dependence of the graviton propagator, which
implies\footnote{The picture is more subtle in the case of an infinite
plasma, and the associated saturation scale shows a faster rise with the
energy, namely $Q_s^2\propto 1/x^2$, but the additional power of $1/x$ is
understood simply as the coherence length of the virtual photon
\cite{HIM2,HIM3}.} $Q_s^2\propto 1/x$. This specific energy dependence is
also necessary to ensure energy--momentum conservation, as we shall later
explain. On the other hand, Ref.~\cite{Albacete:2008ze} reported a rather
counterintuitive result for $Q_s$, which becomes independent of $x$ in
the high energy limit. Although that analysis uses a different projectile
--- namely, a small `color dipole' (dual to a Nambu--Goto open string in
AdS$_5$) instead of a virtual photon ---, it is unlikely that this can
explain the dramatically different result obtained there for the
saturation momentum\footnote{In fact, we shall find that the supergravity
field dual to the virtual photon describes a dipolar partonic fluctuation
of the latter, so the DIS proceeds via the dipole scattering off the
shockwave, so like in QCD at weak coupling. Hence even the physical setup
looks {\em a priori} similar in Refs.~\cite{HIM1,HIM2,Mueller:2008bt} and
respectively \cite{Albacete:2008ze}.}. Indeed, the latter is an intrinsic
property of the target, hence it must be insensitive to the precise
nature of the projectile.

Our analysis in Sects.~3 and 4 will extend the previous analysis
\cite{Mueller:2008bt} of a finite--width plasma, with results which agree
with Ref.~\cite{Mueller:2008bt} whenever a direct comparison is possible.
Our generalization will refer to several aspects:

\texttt{(i)} We shall consider two types of external probes: an
$\mcal{R}$--current (so like in Ref.~\cite{Mueller:2008bt}) in Sect.~4,
and a `dilaton' in Sect.~3. The $\mcal{R}$--current is an analog of the
electromagnetic current for the ${\mathcal N}=4$ SYM theory, and is dual
to an Abelian vector field propagating in the asymptotically AdS$_5$
space--time. In the supergravity approximation, DIS amounts to solving
the Maxwell equations for this vector field in the background of the
AdS$_5$ shockwave geometry\footnote{For the supergravity approximation to
apply to DIS, the total COM energy squared $s$ must be limited in such a
way not to allow for the excitation of massive string states. The
respective condition can be written as $1/x\ll \sqrt{\lambda}$
\cite{Polchinski:2002jw}, where $x\simeq Q^2/s$ is the Bjorken
variable.}. The `dilaton' is a supergravity scalar field which is dual to
the Lagrangian density $\mcal{L}=(1/4)F^a_{\mu\nu}F_a^{\mu\nu}+\dots$.
The dilaton case will be presented first, and in more detail, since the
corresponding equations of motion --- the Klein--Gordon equations in the
shockwave geometry
--- turn out to be simpler.

\texttt{(ii)} We shall for the first time consider the impact parameter
dependence of the scattering amplitude and of the saturation momentum,
and in particular study the expansion of the black disk with increasing
energy and the emergence of the Froissart bound. We shall perform this
analysis for two types of shockwaves: that introduced in
Ref.~\cite{Gubser:2008pc} and which involves a pointlike source in the
bulk, and the one that we shall propose in Sect.~2 below and which allows
for an arbitrary inhomogeneity in the transverse plane.

\texttt{(iii)} Both for the dilaton and for the $\mcal{R}$--current, we
shall construct the exact supergravity solution in great detail, by
resumming multiple scattering to all orders and showing that this leads
to the eikonalization of the single graviton exchange, as expected for
the scattering off a shockwave. (The eikonalization was also advocated in
Ref.~\cite{Mueller:2008bt}, but without an explicit construction.) This
requires the respective `bulk--to--bulk' propagators in AdS$_5$ in real
time and light--cone coordinates, that will be constructed in Appendix \ref{app-G}. Similar results in the AdS/CFT context have been previously obtained in Refs.~\cite{Cornalba:2007zb,Brower:2007xg,Levin:2009vj}.

Let us now summarize the main results and conclusions which will emerge
from our analysis:

The DIS structure functions in the supergravity approximation come out in
a factorized structure which is reminiscent of the
$k_\perp$--factorization in QCD at weak coupling \cite{CGCreviews}, with
the radial distance $z$ in AdS$_5$ playing the role of the transverse
size $r_\perp\sim 1/k_\perp$ of the partonic fluctuations of the
projectile (`virtual photon') in the boundary gauge theory (as expected
from the UV/IR correspondence \cite{Brodsky:2008pg,HIM2,HIM3}). Namely,
they involve a convolution in $z$ over the square of the incoming field
(the supergravity solution for the dilaton or the Maxwell field in the
absence of the shockwave) times the cross--section for the scattering
between this field and the shockwave at a given value of $z$. In turn,
this cross--section is expressed in the eikonal approximation, as the
integral over all the impact parameters of a scattering amplitude which
resums multiple graviton exchanges to all orders.

For sufficiently large $Q^2$ and/or large values of Bjorken $x$, the
multiple scattering series can be expanded out, and then the dominant
contribution to the structure functions comes from the double graviton
exchange (the single graviton exchange being purely real). This leading
contribution is of higher twist order, meaning that there are no
point--like constituents in the target, in agreement with previous
analyses at strong coupling
\cite{Polchinski:2002jw,HIM1,BallonBayona:2007qr,BallonBayona:2007rs,HIM2,HIM3},
but in sharp contrast with the situation at weak coupling, where the
dominant contribution at large $Q^2$ is of twist--two order and describes
parton (in QCD, quark) distributions \cite{Peskin,CGCreviews}.

For sufficiently low $Q^2$ at a given value of $x$, the amplitude reaches
the unitarity limit and the structure functions are large. Remarkably, at
least for the $\mcal{R}$--current, the structure functions at low $Q^2$
have the same parametric form as for a proton in QCD at weak coupling.
This similarity suggests a physical interpretation in terms of parton
saturation at strong coupling \cite{HIM1,HIM2} : the low--$k_\perp$ and
low--$x$ region of the phase--space is filled with partons, with
occupation numbers of order one. Note that, in pQCD, the occupation
numbers at saturation are of order one only for quarks, but they are of
order $1/\lambda$ (with $\lambda=g^2N_c\ll 1$ at weak coupling) for
gluons \cite{CGCreviews}. In that case, saturation is driven by gluon
dynamics, namely by the enhanced radiation of gluons with small values of
$x$ and their mutual interactions (quark saturate only due to their
coupling to gluons). By contrast, at strong coupling, saturation is
driven by quasi--democratic branching and the occupation numbers at
saturation are of order one for all types of partons \cite{HIM1,HIM2}.

The borderline between the weak--scattering regime at high--$Q^2$ and
large--$x$, where there are no partons, and the saturation region at
low--$Q^2$ and low--$x$ defines the {\em saturation line}, which can be
expressed either as $Q^2=Q^2_s(x)$, or $x=x_s(Q)$. We find that this
saturation line is exactly the same for the dilaton and the
$\mcal{R}$--current, which confirms that this is an intrinsic property of
the `nuclear' target. Specifically, we find $Q_s^2(x)\sim \Lambda^3 L/x$,
in agreement with Ref.~\cite{Mueller:2008bt}; here, $\Lambda$ is the
`confinement' scale, as introduced by the energy density in the
target\footnote{This is also the scale which fixes the radial cutoff at
$z\sim 1/\Lambda$ in AdS$_5$, as previously explained, but this cutoff
plays no role for DIS at $Q^2\gg\Lambda^2$.}, and $L$ is the longitudinal
extent of the target in its rest frame. This peculiar $1/x$--dependence
of $Q_s$, which reflects the energy--dependence of the single graviton
exchange, can be also understood via an independent argument, that we now
explain.

Namely, energy--momentum conservation implies that the integral $\int_0^1
\rmd x\,F_2(x,Q^2)$ of the structure function $F_2$ has a finite limit,
of order $N_c^2$, as $Q^2\to\infty$. At large $x\gg x_s(Q)$, $F_2(x,Q^2)$
has only higher--twist contributions which rapidly die away with
increasing $Q^2$ (see \eqnum{Fhigh}). At $x\lesssim x_s(Q)$, $F_2(x,Q^2)$
is independent of $x$, due to saturation, and of order $N_c^2 Q^2$ (see
\eqnum{Flow}). Hence, the integral is dominated by $x\sim x_s(Q)$ --- the
highest value of $x$ at which the structure function is still finite at
large $Q^2$
---, and can be estimated as
 \beq\label{sr1}
 \int_0^1 \rmd x\,F_2(x,Q^2)\,\sim\, x_s F_2(x_s,Q^2)\,\sim\,
 x_s\,N_c^2 Q^2\,.\eeq
For this to be independent of $Q^2$, $x_s$ must scale as $x_s(Q)\sim
1/Q^2$, or $Q^2_s(x)\sim1/x$, as announced.

So far, we did not mention the dependence of the various results, so like
$Q_s$, upon the impact parameter $b_\perp$. This will be discussed at
length in the main text, and the main conclusion is that, at strong
coupling, the function $Q^2_s(x,b_\perp)$ has the same
$b_\perp$--dependence as the energy--momentum distribution in the
shockwave. In particular, for a distribution which exhibits an
exponential tail at large $b_\perp$ (as expected in the presence of
confinement), $Q^2_s(x,b_\perp)$ has an exponential tail as well, and the
black disk area grows like $\ln^2(1/x)$, that is, it saturates the Froissart
bound. On the other hand, for the shockwave metric in
Ref.~\cite{Gubser:2008pc}, $Q^2_s(x,b_\perp)\sim 1/b_\perp^{\,6}$ at
large $b_\perp$, and the area of the black disk grows like a power of the
energy, in violation of the Froissart bound.

\section{Shockwaves with impact parameter dependence in AdS$_5$}
\label{sect-SW} \setcounter{equation}{0}

In this section, we shall discuss various proposals for asymptotically
AdS$_5$ shockwave metrics which are intended to represent the gravity
duals of a fast moving `nucleus' (more properly, a slice of the plasma)
with a non--trivial profile in impact parameter space (the two
dimensional space transverse to the direction of motion). We shall work
in the nucleus infinite momentum frame, that is, we shall take the
nucleus to move in the positive $x^3$ direction with a Lorentz $\gamma$
factor which is arbitrarily large. It is then convenient to introduce
light--cone coordinates,
 \beq
 x^+\,\equiv\,\frac{1}{\sqrt{2}}\,(x^0+x^3)\,,\qquad
 x^-\,\equiv\,\frac{1}{\sqrt{2}}\,(x^0-x^3)\,,\eeq
in terms of which the nucleus moves in the positive $x^+$ direction, and
the only non--trivial component of its average energy--momentum tensor
$\langle T_{\mu\nu}\rangle$ is the $(-,-)$ component (as this is the only
one to be enhanced by the large factor $\gamma^2$). By energy--momentum
conservation, $\del^\mu\langle T_{\mu\nu}\rangle=0$, this
component\footnote{To avoid cumbersome notations, we shall omit the
brackets denoting expectation values whenever there is no risk of
confusion.} $T_{--}\equiv \langle T_{--}\rangle$ is independent of the
light--cone time $x^+$. For the time being, we shall allow
$T_{--}(x^-,x_\perp)$ to be an arbitrary function of $x^-$ and the
transverse coordinates $x_\perp=(x^1,x^2)$ (the `impact parameter'). In
practice we shall be mostly interested in the situation where the nucleus
is strongly Lorentz contracted: $T_{--}\propto \delta(x^-)$ (the
`shockwave').

We shall assume $T_{--}$ to be proportional to $N_c^2$. This is unrealistic from the point of view of a real QCD problem, where a nucleus is built with colorless hadrons (protons and nucleons). But within the present AdS/CFT context, such an assumption is necessary in order to be able to study the high--energy dynamics in the large--$N_c$ limit, to which we would like restrict ourselves in what follows. Indeed, it is only when $T_{--}\propto N_c^2$ that the metric perturbation induced by the `nucleus' in the `bulk' metric of AdS$_5$ is an effect of $\order{1}$ which survives when $N_c\to \infty$. Then, the nucleus and its interactions can be simply described in the supergravity approximation, in which one first solves the Einstein
equations to determine the asymptotically AdS$_5$ metric dual to the
nucleus and then study the propagation of various projectiles in this
metric.

Our subsequent construction may be viewed as a model for the gravity dual
of a slice of a ${\mathcal N}=4$ SYM plasma at finite temperature $T$,
which indeed has $N_c^2$ degrees of freedom per unit volume, and hence an
energy density $T_{00}\sim N_c^2 T^4$ in the plasma rest frame, yielding
$T_{--}\sim \gamma^2 N_c^2 T^4$ in the infinite momentum frame. In
theoretical studies, the plasma is generally assumed to be infinite, but
this is unrealistic from a phenomenological viewpoint and, besides, the
concept of plasma makes sense also for a finite volume system (at least
over a finite time interval), so long as the size $L$ of the system along
any direction obeys $L\gg 1/T$. An explicit connection between the
shockwave metric that we shall construct in this section and the boosted
slice of the plasma will be established in Appendix A.

As mentioned in the Introduction, there are two types of proposals in the
literature for shockwave metrics in AdS$_5$ : one which introduces a
gravity source in the bulk of AdS$_5$
\cite{Gubser:2008pc,AlvarezGaume:2008fx,Lin:2009pn,Gubser:2009sx}, and
one which does not \cite{Janik:2005zt,Beuf:2009mk} (but {\em boundary}
sources are in principle allowed in the second case; see below). Both
cases can be encoded in the following Einstein equations
 \beq\label{Einstein}
\mathcal{R}_{mn}\,- \,\frac{\mathcal{R}}{2}\,g_{mn}
 \,+\Lambda\, g_{mn}\,=\,8 \pi G_5\,J_{mn}\,,\eeq
where $m,\,n$ are 5--dimensional space--time indices, $\Lambda
=-{6}/{R^2}$ (with $R$ the curvature radius of AdS$_5$) is the
cosmological constant, $\mathcal{R} \equiv g^{mn} \mathcal{R}_{mn} =
-20/{R^2}$, $G_5$ is the Newton constant in $D=5$, and $J_{mn}$ is the
stress tensor of the source localized in the bulk, for which we shall
consider the two scenarios alluded to above:

\begin{enumerate}

\item \texttt{Type-I metric:} $J_{mn}=0$. This is the case considered
    in Refs.~\cite{Janik:2005zt,Beuf:2009mk}.

\item \texttt{Type-II metric:} $J_{mn}$ corresponds to an ensemble of
    $N_c^2$ point--like `particles' moving together within AdS$_5$
    along a null geodesic parallel to the Minkowski boundary (see
    \eqnum{Jbulk} for an explicit expression). This is the case
    considered in
    Refs.~\cite{Gubser:2008pc,AlvarezGaume:2008fx,Lin:2009pn,Gubser:2009sx}.

\end{enumerate}

\noindent
As shown in the literature, the corresponding shockwave
solutions to \eqnum{Einstein} can be obtained with the following
\emph{Ansatz} (in the so--called Fefferman--Graham coordinates)
 \beq\label{metric}
\rmd s^2 \,= \,\frac{R^2}{z^2} \,\big[ \rmd z^2 -2 \rmd x^+ \rmd x^- +
\rmd x_\perp^2 + h(z,x^-,x_\perp) (\rmd x^-)^2\big],
 \eeq
for {\em both} cases: with or without bulk sources. Still in both cases,
the function $h(z,x^-,x_\perp)$ is determined by the following, {\em
linear}, equation
 \beq
\Big(\partial_z^2 - \frac{3}{z}\,\partial_z + \nabla_\perp^2\Big) \,
h(z,x^-,x_\perp) \,=\, -16\pi G_5\,J_{--}\,, \label{bequation}
 \eeq
which is the $(-,-)$ component of \eqnum{Einstein}. Note that there is no
`small perturbation' assumption involved here: \eqnum{bequation} is the
exact consequence of the Einstein equations \eqref{Einstein} for the
\emph{Ansatz} \eqref{metric} and the specific stress--tensor $J_{mn}$
under consideration. Accordingly, the metric `perturbation'
$g_{--}=(R^2/z^2)h$, which describes the shockwave, needs not be
parametrically small. In fact, the normalization of this perturbation is
related to that of the energy--momentum tensor on the boundary, via
holographic renormalization
 \cite{deHaro:2000xn,Skenderis:2002wp}. Specifically, if
$h^{(4)}(x^-,x_\perp)$ is the coefficient of the $z^4$ term in the
near--boundary ($z\to 0$) expansion of $h(z,x^-,x_\perp)$, then
 \beq\label{HREN}
T_{--}(x^-,x_\perp)\,= \,\frac{R^3}{4\pi G_5}\,h^{(4)}(x^-,x_\perp)
 \,= \,\frac{N_c^2}{2\pi^2}\,h^{(4)}(x^-,x_\perp)\,,
 \eeq
where the second equality follows after using the AdS/CFT correspondence
to identify $G_5=\pi R^3/2N_c^2$. As anticipated, $T_{--}$ must scale
like $N_c^2$ for the metric perturbation not to be parametrically small.

We shall now successively consider the two interesting cases. In this
process, we shall recall some results from the literature, and we shall
correct and extend the proposal in Refs.~\cite{Janik:2005zt,Beuf:2009mk}.

\subsection{Type--I metric: no source in the bulk}

In this case we need the non--trivial solutions to the homogeneous
version of \eqnum{bequation}. Consider first the case where there is no
dependence upon $x_\perp$, corresponding to a nucleus which is uniform in
impact parameter space (an infinite, uniform, wall). Then the solution
reads
 \beq\label{bhom}
 h(z,x^-)\,=\,\frac{2\pi^2}{N_c^2}\,z^4\,T_{--}(x^-)\,,\eeq
where the normalization has been fixed according to \eqnum{HREN}. With
this expression for $h$, the metric \eqref{metric} is the original
shockwave metric proposed by Janik and Peschanski \cite{Janik:2005zt}.
Although an exact solution to the (homogeneous) Einstein equations, this
metric has nevertheless the drawback to have a singular point at
$z\to\infty$. We shall return to this issue later on.

Consider now the homogeneous version of \eqnum{bequation} corresponding
to a generic profile in $x_\perp$. It is then convenient to perform a
Fourier transform to transverse momentum space, which yields the following equation
 \beq
\Big(\partial_z^2 - \frac{3}{z}\,\partial_z -k_\perp^2\Big) \,
h(z,x^-,k_\perp) = 0. \label{beqk}
 \eeq
The general solution is expressed in terms of modified Bessel functions
of second rank:
 \beq
h(z, x^-, k_\perp) = \frac{1}{2}\,(zk_\perp)^2 \big[c_1(x^-, k_\perp)
\rmK_2(zk_\perp) +  c_2(x^-, k_\perp) \rmI_2(zk_\perp)\big].
\label{bsolution}
\eeq
The near--boundary expansion of this solution reads as follows (up to
order $z^4$)
\beq\label{expbnd}
h(z, x^-, k_\perp)  &=& c_1(x^-, k_\perp)\Big[1 - \frac{z^2k_\perp^2}{4}+
\frac{z^4k_\perp^4}{32}\big(-2\ln(zk_\perp/2) + 3/2 -2\gamma_E\big)\Big]
 + \nonumber \\
&{}&\,+\, c_2(x^-, k_\perp) \,\frac{z^4k_\perp^4}{16} \,  + \cdots\,,
\eeq
that is, the component proportional to $\rmI_2(zk_\perp)$ vanishes like
$z^4$ when $z\to 0$, while that proportional to $\rmK_2(zk_\perp)$
approaches a non--zero value in this limit. Consider also the large $z$
behavior: for $zk_\perp\gg 1$, $\rmK_2(zk_\perp)$ is exponentially
decreasing, while $\rmI_2(zk_\perp)$ is exponentially increasing. For
what follows, it is interesting to keep in mind that the homogeneous (in
the sense of no dependence upon $x_\perp$) solution in \eqnum{bhom}
corresponds to the limit $k_\perp\to 0$ of the $\rmI_2$--piece of the
general solution \eqnum{bsolution}.

In the framework of AdS/CFT, the solutions which diverge as $z\to\infty$
are unacceptable and must be discarded. In view of that, it might look
natural to enforce $c_2=0$ in \eqnum{bsolution}: this is the common
strategy for computing correlation functions in the vacuum of ${\mathcal
N}=4$ SYM (here, the correlators of $T_{\mu\nu}$)
\cite{Maldacena:1997re,Gubser:1998bc,Witten:1998qj}, and this was also
the proposal made in Ref.~\cite{Beuf:2009mk} for constructing the gravity
dual of a nucleus. However, in what follows we shall argue that choosing
$c_2=0$ in \eqnum{bsolution} leads to a physically unacceptable picture
for a nucleus. (In particular, this would also exclude the homogeneous
shockwave \eqref{bhom}, which as alluded to above represents the limit
$k_\perp\to 0$ of the piece of the solution proportional to $c_2$.) A
more sensible choice, which is physically motivated, is to take $c_1=0$
and introduce a cutoff in the radial direction of AdS$_5$ at a distance
$z_\Lambda= 1/\Lambda$, with $\Lambda$ playing the role of an infrared
cutoff in the boundary gauge theory. To motivate this proposal, let us
first explain the difficulties with the original choice in
Ref.~\cite{Beuf:2009mk}.

As just mentioned, taking $c_2=0$ is the standard choice for computing
the correlators of $T_{\mu\nu}$ in ${\mathcal N}=4$ SYM. In that case,
the metric perturbation has a non--zero limit on the boundary,
 \beq
 h_{\mu\nu}(x^-, k_\perp)\,=\,\lim_{z\to 0}\,\left[\frac{z^2}{R^2}\
 g_{\mu\nu}(z,x^-, k_\perp)\right]\,-\,\eta_{\mu\nu}\,=\,\delta_{\mu-}
 \delta_{\nu-}\,c_1(x^-, k_\perp)\,,\eeq
which acts as an external source (not to be confused with the bulk source
$J_{mn}$ in \eqnum{Einstein}), which couples to $T_{\mu\nu}$ in the
boundary gauge theory. This external source induces a non--vanishing
expectation value for $T_{\mu\nu}$, which is proportional to the source
and can be read off \eqnum{HREN} :
 \beq
\langle T_{\mu\nu}(x^-, k_\perp)\rangle =\, \delta_{\mu-}
 \delta_{\nu-} \,\frac{N_c^2
}{64\pi^2}\,k_\perp^4\ln\left( \frac{\mu^2}{k_\perp^2}\right)\,c_1(x^-,
k_\perp)\,. \label{EMtensor}
\eeq
Here, $\mu$ is scale for ultraviolet renormalization in the gauge theory,
as introduced by the removal of the logarithmic singularity at $z\to 0$
manifest in \eqnum{expbnd}. (The finite terms beyond the logarithm which
are also visible in \eqnum{expbnd} have been absorbed in the definition
of $\mu$.) By taking a functional derivative in \eqnum{EMtensor} with
respect to $h_{--}=c_1$, one obtains the only non--vanishing 2--point
function of $T_{\mu\nu}$ within the present kinematics (in momentum space
and with trivial delta functions removed) :
 \beq
\langle T_{--}(k_\perp)\, T_{++}(-k_\perp)\rangle =\,\frac{N_c^2
}{64\pi^2}\,k_\perp^4 \ln\left( \frac{\mu^2}{k_\perp^2}\right)\,.
\label{EM2point}
\eeq
This is indeed the expected result\footnote{In the corresponding
expressions in the literature, $k_\perp^2$ is generally replaced by the
invariant 4--momentum squared $k^2=k_\perp^2-2k^+k^-$. Note however that
for the present set--up we have $k^-=0$ since there is no dependence upon
$x^+$.}, with the specific $k_\perp$--dependence in the r.h.s. reflecting
the conformal symmetry of ${\mathcal N}=4$ SYM (see, e.g., the discussion
in \cite{Gubser:1998bc}).

Within this standard procedure, the external source $h_{\mu\nu}$ plays no
dynamical role, but rather acts as a device for generating {\em vacuum}
correlations via functional differentiation. By contrast, in
Ref.~\cite{Beuf:2009mk} it has been proposed to use a similar procedure
with a suitable choice for the function $c_1(x^-, k_\perp)$ in such a way
to generate, via \eqnum{EMtensor}, an energy--momentum profile which
physically would correspond to an ultrarelativistic nucleus. However,
Ref.~\cite{Beuf:2009mk} showed no explicit proposal for such a function
$c_1(x^-, k_\perp)$, and in fact we shall now argue that there is no
meaningful solution of this type.

Our main objection to the proposal in Ref.~\cite{Beuf:2009mk} is of
physical nature, but it also gets reflected in severe technical
difficulties, which lead to paradoxes. Let us start with one of these
paradoxes, before we explain the deeper origin of the problem. Returning
to the simpler notation $T_{--}\equiv \langle T_{--}\rangle$, we notice
that one must have $T_{--}(k_\perp = 0) > 0$ because
\beq
T_{--}(x^-, k_\perp = 0)\, = \int \rmd ^2x_\perp\, T_{--}(x^-, x_\perp)
\eeq
where $T_{--}(x^-,x_\perp)$ is the positive semidefinite energy density
in the transverse plane. For this to be consistent with \eqnum{EMtensor}
one should have $c_1$ behaving like $1/k_\perp^4$ when $k_\perp \to 0$.
But then the Fourier transform of $h(z, x^-, k_\perp)$ back to the
transverse coordinate space is ill defined, because of an infrared
divergence at $k_\perp \to 0$. On the other hand, there is clearly no
similar problem if one chooses $c_1=0$ and $c_2\ne 0$ in
\eqnum{bsolution}, since then one can take $c_2\sim 1/k_\perp^4$ as
$k_\perp \to 0$ without generating infrared problems, as obvious from the
expansion \eqref{expbnd}.

That this is the only meaningful choice, it is also suggested by the
behavior \eqref{expbnd} of the metric near $z=0$ together with its
physical interpretation according to the AdS/CFT dictionary. Via the
UV/IR correspondence
\cite{Susskind:1998dq,Peet:1998wn,Polchinski:2002jw,Brodsky:2008pg,HIM2,HIM3},
the inverse $1/z$ of the radial distance in AdS$_5$ is mapped onto the
transverse momenta (or virtualities) of the quantum fluctuations in the
boundary gauge theory. Hence an expression like \eqnum{bsolution} for the
metric perturbation should be viewed as encoding information about {\em
two} types of transverse momenta: the momentum $k_\perp$ which via the
Fourier mode $c_2(x^-, k_\perp)$ determines the profile of the nucleus in
impact parameter space, and the momentum $p_\perp\sim 1/z$ which refers
to the quantum modes that we would like to include in our description of
the nucleus and of its high--energy interactions.

On physical grounds, in particular in view of our experience with QCD, we
expect the bulk of the modes in a hadron wavefunction to be concentrated
at `soft' momenta, of the order of some infrared cutoff $\Lambda$,
whereas at much larger momenta $p_\perp\gg \Lambda$ the distribution
should rapidly decrease, typically according to a power law
$(1/p_\perp)^\Delta$. The exponent $\Delta$ is equal to 2 for the
partonic tail produced via bremsstrahlung in QCD at weak coupling, but it
is equal to 4 or larger in all the situations where a gravity dual has
been identified for a hadronic system at strong coupling. For instance,
$\Delta=4$ in the case of the AdS$_5$ black--brane geometry dual to the
${\mathcal N}=4$ SYM plasma \cite{Witten:1998zw}, and also for the
uniform shockwave metric in \eqnum{bhom} \cite{Janik:2005zt}, or for the
metric induced by a pointlike source in the bulk of AdS$_5$
\cite{Gubser:2009sx} (see \eqnum{bgsmallz} below). Furthermore, for a
normalizable dilaton state in AdS$_5$ \cite{Gubser:1998bc,Witten:1998qj},
which is dual to a `glueball', one has $\Delta=2+\sqrt{m^2R^2+4}\ge 4$,
where the dilaton mass $m$ is proportional to the scale $\Lambda$ which
enters via the radial cutoff at $z\sim 1/\Lambda$
\cite{Polchinski:2001tt}. Such a large value for $\Delta$ at strong
coupling can be understood as the absence of partons with high momenta
\cite{Polchinski:2002jw,HIM1,HIM2,HIM3}, an interpretation to which we
shall return in Sects.~3 and 4. Clearly, $\Delta=4$ is also the
prediction of \eqnum{bsolution} provided one takes $c_1=0$ (cf.
\eqnum{expbnd}). On the other hand, with $c_2=0$, \eqnum{bsolution}
predicts a dominant behavior near $z=0$ which is independent of $z$,
that is, a flat momentum spectrum ($\Delta=0$) in the high momentum
regime, which physically makes no sense.

The previous examples also show that the behavior in $z^4$ of the metric
perturbation holds only for sufficiently small values of $z$,
corresponding to the ultraviolet behavior of the dual gauge theory. But
this growth is cut off at some larger value of $z$, whose inverse plays
the role of an infrared cutoff in the gauge theory. Such a cutoff is
necessary to introduce the analog of {\em matter} (plasma or hadronic
bound state) in the otherwise conformal SYM field theory. For instance,
this scale is provided by the black hole horizon at $z\sim 1/T$ for the
AdS$_5$--Schwarzschild metric, by the radial position of the source in
the bulk for the shockwave metric in Ref.~\cite{Gubser:2009sx} (see
Sect.~\ref{Sect-TypeII} below), and by an explicit cutoff at large values
of $z$ in the construction of the glueball bound state
\cite{Polchinski:2000uf,Polchinski:2001tt,Karch:2006pv}.

At a first sight, the uniform shockwave metric \eqref{bhom} seems not to
involve any such a scale, but this is only illusory: written as it
stands, the metric perturbation in \eqnum{bhom} blows up at $z\to\infty$
and this divergence is a genuine space--time singularity, and not just a
coordinate singularity. As mentioned in the Introduction, this
singularity is quite elusive, as it does not show up in any of the
curvature scalars computed from this metric. Rather, it can be identified
via a more general criterion for space--time singularities, known as
$g$--incompleteness (with $g$ standing for ``geodesic'') \cite{Hawking}.
Namely, in an acceptable space--time,  which is $g$--complete, any
time--like or null--like geodesic can be extended up to arbitrary values
in their affine parameters (like the proper time). The contrary would
imply the existence of observers whose history ends or begin at a finite
proper time. On the other hand, for the Janik--Peschanski metric one can
construct time--like geodesics which, starting at a generic point $z_0$,
reach infinity after a finite value of the proper time. (See
\cite{Podolsky:1997ik} for an analogous construction in the Kaigodorov
space--time, which is the 4--dimensional version of the Janik--Peschanski
metric.) Hence, this metric is $g$--incomplete, and thus unacceptable.
One can effectively `hide' the space--time singularity at $z\to\infty$ by
introducing a radial cutoff at $z=1/\Lambda$, which for that purpose
plays the same role as the black hole horizon within the AdS$_5$
black--brane metric. But once this is done, it is no more disturbing to
use a general metric perturbation $\propto z^2\rmI_2(zk_\perp)$, which
has the right behavior at small $z$ and allows for a generic transverse
inhomogeneity, instead of the uniform shockwave in \eqnum{bhom}.

To summarize, our proposal for a gravity dual to a large nucleus is given
by the asymptotically AdS$_5$ metric in \eqnum{metric} which applies for
$z \le z_\Lambda\equiv 1/\Lambda$ and where the function $h$ is given, in
transverse momentum space, by \eqnum{bsolution} with $c_1=0$ and $c_2$
related to the energy--momentum tensor of the nucleus that we would like
to describe via
\beq
T_{--}(x^-, k_\perp)\,= \,\,\frac{N_c^2 }{32\pi^2}\,k_\perp^4\, c_2(x^-,
k_\perp). \label{EMtensor2}
\eeq
The example of the `plasma slice' previously discussed suggests that a
physically reasonable choice for $T_{--}$ would be
 \beq\label{Tmmsimple}
  T_{--}(x^-,x_\perp) = \gamma^2 N_c^2 \Lambda^4 f(x^-,x_\perp\Lambda)\,,
 \eeq
where the dimensionless function $f$ describes the shape of the `nucleus'
in longitudinal and transverse directions. The transverse inhomogeneity
is controlled by the `soft' scale $\Lambda$, so like in QCD. The
longitudinal support is concentrated at $|x^-|\lesssim L/\gamma$ with $L$
the width of the `nucleus' in its rest frame. In the high energy limit,
we can replace this by a $\delta$--function in $x^-$ :
 \beq\label{fprof}
 f(x^-,x_\perp\Lambda)\,=\,f(x_\perp\Lambda)\,
 \frac{L}{\gamma}\,\delta(x^-)\,,
 \qquad \Lambda^2\int \dif^2 x_\perp\,  f(x_\perp\Lambda)\, =\,1\,,
 \eeq
where the normalization of the new function $f(x_\perp\Lambda)$ has been
chosen for convenience. Some reasonable choices for $f$, inspired by our
experience with QCD, would be a Gaussian, or a Woods--Saxon profile which
falls exponentially for distances far away from the center, that is for
$x_{\perp}\Lambda \gg 1$. These choices lead to the following model for
the metric perturbation:
\beq\label{bl}
h(z, x^-, x_\perp) \,=\,16\pi^2\gamma L \Lambda^2 z^2\,\delta(x^-)\int
\frac{\rmd^2 k_\perp}{(2\pi)^2k_\perp^2} \  \rme^{ik_\perp x_\perp} \,
\tilde f(k_\perp/\Lambda)\, \rmI_2(z\,k_\perp)\,,
\eeq
where $\tilde f(k_\perp/\Lambda)$ is the Fourier transform of $\Lambda^2
f(x_\perp\Lambda)$ (so it is dimensionless), and it has support at
$k_\perp\lesssim \Lambda$. A radial cutoff at $z\sim z_\Lambda$ is
implicit. Then, clearly, the argument of $\rmI_2$ can never become large.

Our prescription for cutting off AdS at $z= z_\Lambda$ is merely a model,
and any calculation which is sensitive to large values of $z$ (so like
the scattering between two shockwaves) will be strongly sensitive to the
details of this model
--- the value of $\Lambda$ and the specific procedure used for its
implementation. Fortunately, there are also interesting phenomena, so
like the deep inelastic scattering to be considered in Sects.~3 and 4,
which are controlled by the `hard' ($p_\perp\gg \Lambda$) part of the
spectrum --- in the AdS framework, by the behavior of the metric near the
boundary at $z=0$ --- and thus are completely insensitive to the model
used to cutoff AdS$_5$ at large $z$.

\subsection{Type--II metric: pointlike source in the bulk}
\label{Sect-TypeII}

We now briefly describe the shockwave proposal in
Ref.~\cite{Gubser:2009sx}, which involves a source in the bulk. We shall
take this source to be composed of $N_c^2$ point--like `particles' moving
together\footnote{Ref.~\cite{Gubser:2009sx} mentioned only a single such
a particle, but here we shall consider a collection of $N_c^2$ of them,
in order to achieve a metric perturbation of order one. The alternative
possibility, which would be to take a single particle but with ultrahigh
energy $p^+\sim N_c^2$, would be inconsistent with the use of the
supergravity approximation for any collision involving that `particle'
\cite{Polchinski:2002jw,HIM1}.} along the trajectory defined by: $x^-=0$,
$x_\perp=0$, and $z=z_*$, with $z_*$ a positive constant. Then the only
non--zero component of $J_{mn}$ is $J_{--}$, and is given by
\cite{Gubser:2009sx}
 \beq\label{Jbulk}
 J_{--}\,=\, p^+ N_c^2\,\frac{z^3}{R^3}\
 \delta(x^-)\delta^{(2)}(x_\perp)\delta(z-z_*)\,,\eeq
with $p^+$ the light--cone longitudinal momentum of a single `particle'.
The solution to \eqnum{bequation} corresponding to this current reads
 \beq\label{bgubser}
{h}(z,x^-,x_\perp) \,=\, \frac{\pi p^+}{16} \frac{zz_* }{q^3}\,
{}_2F_1(3,5/2,5,-1/q)\, \delta(x^-)
\eeq
where
\beq
q\,\equiv\, \frac{x_\perp^2 + (z-z_*)^2}{4zz_*} \label{chordal}
\eeq
is the so called `chordal distance'\footnote{The chordal distance is the
$SO(3,1)$--invariant distance, \emph{i.e.} the analog of the radial
distance, for the hyperbolic space $H_3$ spanned by the coordinates
$(z,x_\perp)$.} between the point $(z,x_\perp)$ where we measure the
metric and the position $(z_*,x_\perp=0)$ of the pointlike source in the
bulk, and the hypergeometric function takes a rather simple form:
 \beq
 {}_2F_1(3,5/2,5,-1/q)\,=\,
 16 q^3 \frac{1 + 8 q(1 + q)  - 4\sqrt{q(1 + q)}(1+2q)}
 {\sqrt{q(1 + q)}}\,.\eeq
As anticipated, the energy $E$ of the particle must scale like $N_c^2$ in
order for the metric perturbation \eqnum{bgubser} to be non--negligible.
Note that $q\gg 1$ for both small ($z\ll z_*$) and large ($z\gg z_*$)
values of $z$. Using ${}_2F_1\simeq 1-3/2q \approx 1$ when $q\gg 1$, we
deduce
 \beq\label{bgsmallz}
{h}(z,x^-,x_\perp) \,\simeq\, \frac{4\pi p^+(zz_*)^4
 }{(x_\perp^2 + z_*^2)^3}\, \delta(x^-)\qquad\mbox{for}\qquad z\ll z_*
 \,,\eeq
and respectively
 \beq\label{bglargez}
{h}(z,x^-,x_\perp) \,\simeq\,\frac{4\pi p^+(zz_*)^4
 }{(x_\perp^2 + z^2)^3}\, \delta(x^-)\qquad\mbox{for}\qquad z\gg z_*\,,\eeq
so, in particular, the metric perturbation dies away as $1/z^2$ when
$z\to\infty$.

The associated energy--momentum tensor in the boundary gauge theory
follows from \eqnum{HREN}:
 \beq\label{Tgubser}
T_{--}(x^-,x_\perp)
 \,= \,\frac{p^+ N_c^2}{\pi}\,\frac{2z_*^4
 }{(x_\perp^2 + z_*^2)^3}\, \delta(x^-)\,.
 \eeq
This is essentially uniform so long as $x_\perp\ll z_*$ but it decreases
like $1/x_\perp^6$ for $x_\perp\gg z_*$. We see that $1/z^*$ plays the
same role as the `soft' momentum scale $\Lambda$ introduced previously,
in the sense of fixing the scale for transverse inhomogeneity in the
nucleus. But unlike in that previous case, now there is no need to
explicitly cut off the radial dimension of AdS$_5$, since the metric
perturbation \eqnum{bgubser} dies away, like $1/z^2$, when $z\to\infty$.

\section{DIS off the shockwave: the dilaton}
\label{sect-D}

In this and the next coming section, we shall consider the supergravity
problem dual to the deep inelastic scattering (DIS) between a `hard'
(\emph{i.e.}, highly virtual) external current and a nucleus in the
${\mathcal N}=4$ SYM theory. The `nucleus' should be thought off as a
slice of a plasma and it will be described as a shockwave, as already
explained. Its construction requires an infrared cutoff $\Lambda$, but
the details of this cutoff will be unimportant for the hard process at
hand. The `current' generally associated with DIS is a virtual photon
with space--like virtuality which couples to the electromagnetic current
of the nucleus. Within ${\mathcal N}=4$ SYM, this `electromagnetic
current' can be represented by the conserved ${\mathcal R}$--current
carried by (adjoint) fermionic and scalar fields in the Lagrangian. The
DIS of such an ${\mathcal R}$--current will be addressed in
Sect.~\ref{sect-R}. But before doing that, it is preferable to introduce
the formalism in the simpler context of the scalar `current'
$J=(1/4)F^a_{\mu\nu}F_a^{\mu\nu}$ ($a$ is the SU$(N_c)$ color index).
This operator is interesting in itself, since it couples to the gluons
inside the nuclear wavefunction, and hence is a direct probe of the gluon
distribution. The `bulk' AdS field dual to this operator is the massless
dilaton field, which obeys the simplest equation of motion in
supergravity: the Klein--Gordon equation in the relevant (here, the
shockwave) metric. For simplicity, we shall refer to the projectile as
the `dilaton' although, strictly speaking, it is the operator $J$, and
not the dilaton, which undergoes DIS in the boundary gauge theory.

\subsection{Formalism and structure function}

As before, we shall work in the infinite momentum frame of the `target'
(the nucleus), which therefore will be taken to be Lorentz contracted to
a $\delta$--function at $x^-=0$. The `projectile' (the dilaton)
propagates in the negative $x^3$ direction (a `left mover'), with
space--like momentum $q^\mu$. In light--cone coordinates, we have $q^- >
0$ and $q^+ < 0$, and we take $q_\perp=0$ for convenience; hence, the
virtuality of the projectile reads $Q^2 \equiv q^\mu q_\mu= -2q^+q^-
> 0$. As usual with DIS, it is convenient to express the total
cross--section (or the `structure function') in terms of $Q^2$ and the
Bjorken--$x$ variable, defined as
\beq\label{Bj}
x \,\equiv\, \frac{Q^2}{-2q\cdot p} = \frac{Q^2}{2q^- p^+} =
\frac{Q^2}{2q^- \gamma \Lambda}
\eeq
where $p^\mu=\delta^{\mu+}p^+$ is the momentum of a typical constituent
of the target: $p^+ = \gamma \Lambda$ with $\Lambda$ the characteristic
momentum scale in the target rest frame (e.g., $\Lambda=T$ for a slice of
a finite--temperature plasma) and $\gamma\gg 1$ the nuclear boost factor.
The kinematic conditions of interest are such that $Q^2\gg \Lambda^2$ and
$x\ll 1$.

Via the optical theorem, the DIS structure function $F(x,Q^2)$ is
expressed as the imaginary part of the forward scattering amplitude for
Compton scattering:
\beq\label{Fdef}
 F(x,Q^2) = \mathrm{Im} \,\Pi(x,Q^2)\,,
 \eeq
with the `polarization function' $\Pi(x,Q^2)$ usually written as
 \beq\label{JJDIS}
 \Pi(x,Q^2)\,\equiv i\int \rmd^4r\,\rme^{-iq\cdot r}\,
 \langle P\,|\mbox{T}\{J(x) J(y)\}|\,P\rangle\,,\eeq
where $r^\mu=x^\mu-y^\mu$. However, as above indicated, this writing is
appropriate only when the target is in a state $|\,P\rangle$ with given
4--momentum $P^\mu$, which is translationally--invariant. Here, however,
the target is represented as a shockwave which is localized in space and
inhomogeneous in both longitudinal ($x^-$) and transverse ($x_\perp$)
directions. Accordingly, the current--current correlator computed in this
shockwave background depends not only upon the relative separation
$r^\mu$, but also upon the central coordinates $b^-$ and $b_\perp$
(defined as $b^\mu=(x^\mu+y^\mu)/2$). To obtain a structure function
which depends only upon the kinematical variables $x$ and $Q^2$, we shall
follow the prescription in Ref.~\cite{McLerran:1998nk} and average out
the central coordinates. The longitudinal extent of the target is of
order $L/\gamma$ (recall that $L$ denotes its width in its rest frame)
whereas the scale for inhomogeneity in the transverse plane is set by
$\Lambda$. We shall thus replace \eqnum{JJDIS} by (the precise
normalization is irrelevant at this point)
 \beq\label{PI}
\Pi(x,Q^2) &=& (\gamma/L) \Lambda^2\int \rmd^4r \, \rmd b^- \,
\rmd^2b_\perp \, \rme^{iq\cdot r} \,i\Theta(x_0)\, \langle
[J(b+r/2),J(b-r/2)]\rangle\,,
 \eeq
where the expectation value is now computed according to the AdS/CFT
correspondence (see below) and we have replaced for convenience the
time--ordered correlator by the retarded one. Both types of correlators
yield the same imaginary part, but the retarded one is more easy to
evaluate in the context of AdS/CFT; see e.g. \cite{Son:2007vk}. Note that
$J\propto (F_{\mu\nu})^2$ has mass dimension 4, so $\Pi$ will have
dimension 4.

Let $\Pi(x,y)$ denote the (retarded) 2--point function of $J$ which
enters \eqnum{PI}. According to the AdS/CFT correspondence, this is
formally obtained as
\beq
\Pi(x,y)\,=\, \frac{\delta S_{cl}}{\delta \phi_b(x)\delta \phi_b(y)}
\label{GSCL}
\eeq
where $S_{cl}$ is the action of the dilaton field, that is,
\beq\label{SCL}
S \,=\, -\frac{N_c^2}{16 \pi^2R^3} \int \rmd^4x \,\rmd z \,\sqrt{-g}
g^{nm}\partial_n \phi\partial_m \phi\,,
\eeq
evaluated with the solution $\phi(z,x)$ to the classical equations of
motion,
 \beq\label{KG}
\partial_m ( \sqrt{-g}\, g^{mn}\partial_n \phi) = 0\,,
\label{eom}
\eeq
obeying the following boundary condition at $z=0$ :
 \beq\label{BC}
\lim_{z \to 0} \phi (z,x)\,=\,\phi_b(x)\,.
\eeq
On physical grounds, we need the boundary field $\phi_b(x)$ to be the
plane--wave
 \beq\label{PW}
 \phi_b(x)\,= \,\rme^{-i(q^-x^+ +q^+x^-)}\,\tilde\phi_b\,
 \eeq
($\tilde\phi_b$ is simply a number), but at intermediate steps we shall
consider a generic function $\phi_b(x)$, for the purposes of functional
differentiation. Using the equations of motion, one can perform the
integral over $z$ in \eqnum{SCL} and obtain
\beq
S_{cl}= \frac{N_c^2}{16 \pi^2} \int \rmd^4x \,\frac{1}{z^3} \,\phi(z,x)
\,
\partial_z \phi(z,x) \big |_{z=0}  \label{action}
\eeq
Note that the boundary condition \eqref{BC} is not enough to fully
specify the classical solution, since \eqnum{KG} is a second order
differential equation in $z$. Furthermore, the above procedure cannot
reproduce the imaginary part of the retarded correlator, as obvious from
the fact that the 2--point function generated via \eqnum{GSCL} is
symmetric in its arguments. We shall return to these problems later on.

\eqnum{KG} is written in the shockwave metric \eqref{metric}, which is
independent of the light--cone time $x^+$. Thus, the equations are
homogeneous in $x^+$, meaning that the variable $q^-$ introduced by the
projectile is conserved by the dynamics. It is therefore sufficient to
restrict ourselves to the respective Fourier mode, by writing
\beq
\phi(z,x^+,x^-,x_\perp)\, = \,\rme^{-iq^-x^+}\,\phi(z,x^-,x_\perp).
\eeq
(The function $\phi(z,x^-,x_\perp)$ in the r.h.s. depends upon $q^-$, but
this dependence is kept implicit.) Then the equation of motion is more
explicitly written as (with the notation $x = (x^-, x_\perp)$)
 \beq\label{EOM}
 \Big( \partial_z^2 - \frac{3}{z}\partial_z + 2iq^-\partial_- +
 \nabla_\perp^2\Big)\phi(z,x) = -(q^-)^2 h(z,x) \, \phi(z,x).
 \eeq
We have separated in the right hand side the interaction piece, which
describes the scattering between the dilaton and the shockwave. We shall
correspondingly decompose the total field $\phi$ into its `incoming', or
`vacuum', piece $\phi_0$ and the `scattering' piece $\phi_s$:
\beq
\phi(z,x)\, =\, \phi_0(z,x) + \phi_s(z,x)\,,
\eeq
We construct $\phi_0$ in such a way to satisfy the boundary condition
\eqref{BC}, hence
 \beq\label{phiBC}
 \lim_{z \to 0} \phi_0 (x, z) = \phi_b(x), \qquad \lim_{z \to 0} \phi_s
 (x, z) = 0.
 \eeq
The vacuum problem is formally similar to the source--free version of
\eqnum{bequation} that we have already solved: the equation is
homogeneous in space and time, so the solution corresponding to the
boundary field \eqref{PW} is of the form $\phi_0(z,x) =\rme^{-i
q^+x^-}\phi_0(z)$. The function $\phi_0(z)$ obeys the equation obtained
by replacing $k_\perp^2\to -2q^+q^-=Q^2 >0$ in \eqnum{beqk} with the
boundary condition $\phi_0(z\to 0)=\tilde\phi_b$. Clearly the unique
acceptable solution is (compare to \eqnum{bsolution})
  \beq\label{phi0} \phi_0 (z)\, =\,\frac{1}{2}(Q z)^2\, \rmK_2(Q z)\,
  \tilde\phi_b\,.\eeq
This solution, together with Eqs.~\eqref{GSCL} and \eqref{action}, yields
the vacuum component $\Pi_0(Q^2)$ of the polarization function in a form
entirely similar to \eqnum{EM2point} : 
 \beq\label{Pi0}
 \Pi_0(Q^2)\,\equiv i\int \rmd^4x\,\rme^{-iq\cdot x}\,
 \langle 0 \,|\mbox{T}\{J(x) J(0)\}|\,0\rangle\,
 =\,\frac{N_c^2Q^4}
 {64\pi^2}\,\ln\left( \frac{\mu^2}{Q^2}\right)\,.\eeq

For what follows, it is useful to rewrite \eqnum{phi0} in a more general
way, which features the vacuum boundary--to--bulk propagator $D_0(z,x -
y)$ :
 \beq
 \phi_0(z,x) \,=\,\int
 \rmd^3 y \, D_0(z,x - y) \phi_b(y)\,,\qquad D_0(z,k)\,=\,
 \frac{1}{2}\,(K z)^2\, \rmK_2(K z)\,,
 \label{kzero}
 \eeq
where the momentum--space version of $D_0$ was written for the
space--like kinematics of interest here: $K^2\equiv -2k^+q^-+ k_\perp^2
>0$. Via manipulations to be described in the next section,
we will construct the scattering piece $\phi_s$ in a similar form:
\beq
\phi_s (z,x) \,= \,\int \rmd^3 y \, D(z, x, y) \phi_b(y)\,, \label{phis}
\eeq
where $D(z, x, y)$ is the boundary--to--bulk propagator in the shockwave
metric \eqref{metric} and is inhomogeneous in the spatial coordinates.
Inserting this into \eqnum{action} we obtain the scattering piece of the
classical action as
\beq
S_{cl} - S_0 =  \frac{N_c^2}{16 \pi^2} \int \rmd^3 x \int \rmd^3 y \,
\phi_b(x) \, \phi_b(y) \frac{1}{z^3}\, \partial_z D(x, y, z) \biggl
|_{z=0}\,. \label{action2}
\eeq
When acting on \eqref{action2} with the functional
derivatives\footnote{It is now understood that the functional derivatives
are defined with respect to the three--dimensional field
$\phi_b(x^-,x_\perp)$ and for a fixed value of $q^-$.} in \eqref{GSCL},
we obtain two terms:
\beq\label{2D}
\frac{\delta}{\delta \phi_b(x) \, \delta \phi_b(y)} \int \rmd^3 \acute{x}
\int \rmd^3 \acute{y} \, \phi_b(\acute{x}) \, \phi_b(\acute{y})
\frac{1}{z^3}\, \partial_z
D(\acute{x}, \acute{y}, z) \biggl |_{z=0} \nonumber \\
= \frac{1}{z^3}\, \partial_z \left ( D(x, y, z) + D(y, x, z) \right )
\biggl |_{z=0}\,,
\eeq
which together would provide a symmetric and real expression for the
current--current correlator $\Pi(x,y)$. Clearly, this is not the physical
result that we are interested in. To recover the imaginary part of the
retarded 2--point function, and hence the structure function $F(x,Q^2)$,
we follow the prescription in Refs.~\cite{Son:2002sd,Herzog:2002pc} and
drop the second term in \eqnum{2D}, while multiplying the contribution of
the first term by two:
 \beq\label{PiR}
\Pi_R(x,y)\,=\,\frac{N_c^2}{16 \pi^2} \,\frac{2}{z^3}\, \partial_z D(x,
y, z) \biggl |_{z=0}\,.
\eeq
Note that, as it will become explicit later on, the propagator $D(x, y,
z) $ is retarded with respect to the variable $x^-$, which plays the role
of the light--cone time for the dilaton field (a left--mover). Moreover,
the solution built with this propagator, cf. \eqnum{phis}, is such that
for large times $x^-$ it represents a purely `infalling' wave,
\emph{i.e.} a wave which with increasing $x^-$ propagates towards larger
values of the radial dimension $z$. Thus our above prescription for
keeping $D(x, y, z)$ while discarding $D(y, x, z)$ is indeed the same as
the prescription for keeping the infalling solution alone, as originally
formulated in Refs.~\cite{Son:2002sd,Herzog:2002pc}.

Given $\Pi_R(x,y)$, the polarization function \eqref{PI} is finally
computed as
 \beq\label{PI2}
\Pi(x,Q^2) &=& (\gamma/L) \Lambda^2\int \rmd x^- \rmd y^- \,
\rmd^2x_\perp \rmd^2y_\perp \, \rme^{iq^+(x^--y^-)} \,\Pi_R(x,y)\,,
 \eeq
where as compared to \eqnum{PI} we have removed the Fourier transform
over $x^+$ (since this is automatically performed by working in the
$q^-$--representation) and rewritten the measure as, e.g., $\rmd r^- \rmd
b^- = \rmd x^- \rmd y^-$. One can check on the previous equations that
$\Pi_R(x,y)$ has mass dimension 7, hence $\Pi(x,Q^2)$ has dimension 4, as
it should.

In Sect.~\ref{sect-Qsat}, we shall find that the structure function
$F(x,Q^2) = \mathrm{Im} \,\Pi(x,Q^2)$ obeys the following sum--rule
 \beq\label{SR}
\lim_{Q^2\to\infty} \,\frac{\Lambda^2}{Q^2} \int_0^1 \rmd
x\,x\,F(x,Q^2)\,=\,\frac{\pi}{20}\,N_c^2 \Lambda^4
 \,,
 \eeq
which is recognized as the statement of energy--momentum conservation:
this particular integral of $F(x,Q^2)$ singles out the coefficient in
front of $T_{\mu\nu}$ (the energy--momentum tensor operator) in the
operator product expansion (OPE) of the current--current correlator. The
quantity in the r.h.s. of \eqnum{SR} can be recognized as the nuclear
energy density in its rest frame (up to a normalization factor).

\subsection{Eikonal  scattering}

In what follows we shall construct the scattering field $\phi_s$ by
iterating the interaction piece in the r.h.s. of \eqnum{EOM}. This
amounts to resumming graviton exchanges between the target and the
projectile to all orders. To that aim, we need the integral version of
\eqnum{EOM}, that is,
 \beq
 \phi(x^-,x_\perp,z) = \phi_0(x^-,x_\perp,z) + \int \frac{\rmd
 \acute{z}}{\acute{z}^3}\, \rmd y^- \rmd^2y_\perp
 \!\!\!&&G(z,\acute{z}; x^- - y^-, x_\perp - y_\perp)  \nonumber \\
 \times\!\!\!&&[-(q^-)^2] \, h(\acute{z},y^-,y_\perp)\,\phi(\acute{z},y^-,y_\perp),
 \label{phis2}
 \eeq
where $G(z,\acute{z}; x^- - y^-, x_\perp - y_\perp)$ is the
bulk--to--bulk propagator obeying
 \beq\label{eqG}
 \Big( \partial_z^2 - \frac{3}{z}\,\partial_z + 2iq^-\partial_- +
 \nabla_\perp^2\Big)G(z,\acute{z}; x^- - y^-, x_\perp - y_\perp)\,=\,z^3
 \delta(z-\acute{z})\delta(x^- - y^-)\delta^{(2)}(x_\perp - y_\perp)\,,\nn
 \eeq
to be explicitly constructed in Appendix \ref{app-G}. (The euclidean
version of this propagator is well--known in the literature, and will be
recovered in Appendix \ref{app-G}, but here we rather need its real--time
version in light--cone coordinates and mixed Fourier representation.)
Given the boundary conditions \eqnum{phiBC}, it is clear that $G$ must
vanish at the boundary. Several expressions for this propagator will be
useful in what follows. For instance, the following one
 \beq\label{GR}
 G(z,\acute{z}; x^- -y^-, x_\perp - y_\perp) =
 -\frac{\rmi \Theta(x^--y^-)}{2q^-}
 \!\!\!&&\int \frac{\rmd^2 k_\perp}{(2\pi)^2}\,
 \rme^{\rmi k_\perp \cdot (x_\perp- y_\perp)}
 \int_0^\infty \rmd\omega \, \omega
 \nonumber \\
 \times\!\!\! &&
 z^2 \rmJ_2(\omega z)\, \acute{z}^2 \rmJ_2(\omega \acute{z})\,
 \rme^{-\frac{\rmi (x^--y^-)(\omega^2 + k_\perp^2)}{2q^-}},
\eeq
makes it clear that the propagator vanishes as $z\to 0$ and/or
$\acute{z}\to 0$, and also that it is retarded with respect to the
projectile time variable $x^--y^-$. Furthermore, by using the
completeness relation in \eqnum{complet} for the Bessel functions, one
can show that
 \beq\label{delta}
 G(z,\acute{z}; x^- -y^- \to +0, x_\perp - y_\perp) \,= \,-\frac{\rmi}{2q^-}
 \,z^3\delta(z-\acute{z})\delta^{(2)}(x_\perp-y_\perp)\,.\eeq

We now proceed to formally solve \eqnum{phis2} via iterations. The first
iteration gives
 \beq
 \phi_1(z,x^-,x_\perp) = -(q^-)^2 \int \frac{\rmd
 \acute{z}}{\acute{z}^3}\,\rmd \acute{x}^-\rmd^2 \acute{x}_\perp
 \!\!\!&&G(z,\acute{z}; x^- -\acute{x}^-, x_\perp - \acute{x}_\perp)
 \nonumber \\
 \times \!\!\! && h(\acute{z},\acute{x}^-,\acute{x}_\perp)
 \phi_0(\acute{z},\acute{x}^-,\acute{x}_\perp),
 \eeq
which we shall compactly rewrite as
 \beq
 \phi_1(X) = -(q^-)^2 \int \rmd \acute{X}\, G(X,\acute{X}) \,
 h(\acute{X})\, \phi_0(\acute{X}).
 \eeq
The second iteration gives
 \beq\label{2nd}
 \phi_2(X) = (q^-)^4 \int \rmd \acute{X} \, \dacute{X} \, G(X,\acute{X})\,
 h(\acute{X})\, G(\acute{X},\dacute{X})\, h(\dacute{X}) \,
 \phi_0(\dacute{X}),
\eeq
and so on. An important simplification occurs in the limit where the
shockwave $h$ is treated as a $\delta$--function in $x^-$; then, the
perturbative series produced by iterations exponentiates and yields an
{\em eikonal phase} --- a natural result at high energy. Specifically,
let us write $h(z,x^-,x_\perp) = \delta(x^-)\tilde{h}(z,x_\perp)$. Then
from the factor $h(\acute{X})G(\acute{X},\dacute{X})h(\dacute{X})$ in
\eqnum{2nd} we obtain
 \beq
 \delta(\acute{x}^-)\Theta(\acute{x}^--x^-)\delta(\dacute{x}^-) =
 \,\frac{1}{2} \, \delta(\acute{x}^--\dacute{x}^-)\delta(\acute{x}^-)
 \eeq
which after also using \eqnum{delta} gives a factor
 \beq
 -\frac{\rmi}{2q^-}\,\frac{1}{2}
 \,\delta(\acute{X}-\dacute{X})\delta(\acute{x}^-)\,.
 \eeq
Thus \eqnum{2nd} reduces to
 \beq
 \phi_2(X) = -(q^-)^4\frac{\rmi}{2q^-}
 \int \rmd \acute{X}\,
 \frac{1}{2}\,G(X,\acute{X})\, \delta(\acute{x}^-) \,
 \tilde{h}^2(\acute{X})\, \phi_0(\dacute{X}).
 \eeq
This procedure is easily generalized to higher orders: for the $k$--th
iteration we deduce
 \beq
 \frac{1}{k!}\, [-(q^-)^2]^k \, \tilde{h}^k(\acute{X})\,
 \left (-\frac{\rmi}{2q^-} \right)^{k-1}\,,
 \eeq
where the factorial is again generated via the product of theta functions
in the propagators and delta functions in the metric field $h$. Then the
sum exponentiates, as anticipated, with the following final result:
 \beq\label{phisol}
 \phi(z, x^-, x_\perp) = \phi_0(z, x^-, x_\perp) - 2q^- \int \frac{\rmd
 \acute{z}}{\acute{z}^3}\,\rmd^2 \acute{x}_\perp\, G(z,\acute{z}; x^-,
 x_\perp - \acute{x}_\perp) \,\mathcal{T}(\acute{z},\acute{x}_\perp)\,
 \phi_0(\acute{z},0, \acute{x}_\perp)\nn
 \eeq
where we have defined the scattering amplitude corresponding to a radial
penetration $z$ for the dilaton and in the eikonal approximation as
\beq\label{T}
-i\mathcal{T}(z,b_\perp)\,\equiv\,1\,-\, \exp \left[\frac{\rmi
q^-}{2}\,\tilde{h}(z,b_\perp)\right].
\eeq
An interpretation of this quantity in the boundary gauge theory will be
shortly given, following \eqnum{strucfunc}.

Using the expression of $\phi_s$ in \eqnum{phisol} it is now
straightforward to compute the DIS structure function according to
Eqs.~\eqref{Fdef}, \eqref{PiR} and \eqref{PI2}. First we see that the
bulk--to--boundary propagator defined in equation \eqref{phis} is given
by (cf.~\eqnum{kzero})
 \beq
 D(x, y, z) = -2q^- \int \rmd^2 b_\perp
 \int \frac{\rmd \acute{z}}{\acute{z}^3}\,
 G(z, \acute{z}; x^-, x_\perp-b_\perp)\,
 \mathcal{T}(\acute{z},b_\perp) D_0(\acute{z}, -y^-,b_\perp - y_\perp). \label{eq:K}
 \eeq
Clearly the integrals over $x^-$ and $y^-$ in \eqnum{PI2} select the
Fourier components with $k^+=q^+$ in both $G$ and $D_0$. So we are left
with
 \beq
 \Pi(x,Q^2) =- \frac{\gamma\Lambda^2N_c^2}{16 \pi^2 L}
 \,2q^-\int \rmd^2 x_\perp \rmd^2 y_\perp \int \rmd^2 b_\perp
 \int \frac{\rmd \acute{z}}{\acute{z}^3} \!\!\!&&
 \frac{2}{z^3}\,\partial_z\, G(z,\acute{z};q^+,x_\perp - b_\perp)
 \bigg |_{z=0} \nonumber \\
 \times \!\!\! &&
 \mathcal{T}(\acute{z},b_\perp)D_0(\acute{z},q^+,b_\perp-y_\perp).
 \eeq
Using the transverse momentum space form \eqref{Gspace} for the Green's
function,  we see that the $z$ piece in the above reads
 \beq\label{derz}
 \frac{2}{z^3}\,\partial_z\, G(z,\acute{z};q^+,k_\perp) \biggl |_{z=0}
 \,=\,
 -(K\acute{z})^2\rmK_2(K\acute{z})
\eeq
where $K^2 = k_\perp^2 - 2q^+q^- = k_\perp^2+Q^2$. By also using
\eqnum{kzero} for $D_0$, we deduce
 \beq
 \Pi(x,Q^2) = \frac{\gamma\Lambda^2N_c^2}{32 \pi^2 L} \,2q^-
 \int \rmd^2 x_\perp \rmd^2 y_\perp \int \rmd^2 b_\perp
 \int \frac{\rmd z}{z^3}
 \int \frac{\rmd^2k_\perp}{(2\pi)^2}\
 \rme^{\rmi k_\perp \! \cdot (x_\perp-b_\perp)} \nonumber \\
 (K z)^2\rmK_2(K z)\, \mathcal{T}(z,b_\perp)
 \int \frac{\rmd^2q_\perp}{(2\pi)^2}\,
 \rme^{\rmi q_\perp \!\cdot (b_\perp-y_\perp)}
 (\tilde{Q}z)^2 \rmK_2(\tilde{Q}z),
 \eeq
with $\tilde{Q}^2 = Q^2 + q_\perp^2$. This expression simplifies
drastically after performing the transverse coordinate integrals, which
set $k_\perp=q_\perp=0$. Recalling the definition Eq.~\eqref{Bj} we
finally arrive at
 \beq\label{strucfunc}
 \Pi(x,Q^2) \,= \,\frac{Q^6\Lambda N_c^2}{32 \pi^2 xL}
 \int \rmd z \, z \, \rmK_2^2(Q z)
 \int \rmd^2 b_\perp\,
 \mathcal{T}(z,b_\perp)\,,
\eeq
from which the structure function $F(x,Q^2)$ is finally obtained by
taking the imaginary part.

This result has a natural interpretation: $\rmK_2^2(Q z)$ plays the role
of the wavefunction squared for a partonic fluctuation of the `current'
$J=(1/4)F^2$ which according to the UV/IR correspondence has a transverse
size $r_\perp\sim z$. The modified Bessel function 
effectively restricts $r_\perp$  to values $r_\perp\sim z \lesssim 1/Q$,
as expected from the uncertainty principle. Furthermore,
 \beq\label{sigma}
 \sigma(z,x)\,=\,2\int \rmd^2 b_\perp\,
 {\rm Im}\,\mathcal{T}(z,b_\perp)\,\eeq
is the total cross--section for the scattering between this partonic
fluctuation and the `nucleus'.

Let us finally verify that the use of the $\delta$--function
approximation for the metric perturbation, and hence the eikonal
approximation, are indeed justified for the problem at hand. By
inspection of the previous manipulations, it is clear that the only place
where the assumption that $h\propto\delta(x^-)$ has played a role was in
using
 \beq
 \frac{x^-}{2q^-}\,(\omega^2 + k_\perp^2)\,\ll\,1
 \eeq
in order to simplify the last exponential in \eqnum{GR} and thus replace
the intermediate propagators in the iterative series by
$\delta$--functions, cf.~\eqnum{delta}. In reality $x^-$ is, of course,
not strictly zero, but rather of order $L/\gamma$, which is the
longitudinal width of the target in the infinite momentum frame. Since
$k_\perp$ can never be too large (this is set by the transverse
inhomogeneity in the target, so $k_\perp\sim \Lambda$), while $\omega$ is
typically of order ${1/z}$ (as fixed by the oscillatory behavior of the
Bessel functions in \eqnum{GR}), the above inequality amounts to
 \beq
 \label{largez} z^2 \,\gg\,
 \frac{L}{2q^-\gamma}\,=\,\frac{x\Lambda L}{Q^2}\,.
 \eeq
Note that, in this argument, $z$ is the radial distance at some generic
scattering point, e.g.~$\acute{z}$ or $\dacute{z}$ in \eqnum{2nd}, and
not the argument $z$ of $\phi$ in \eqnum{phisol}, which approaches zero
when computing the structure function, cf. \eqnum{derz}. As we shall
shortly check, the integral over $z$ in \eqnum{strucfunc} is dominated by
$z^2\sim 1/Q^2$\,; hence, the condition in \eqnum{largez} is well
satisfied whenever $x\Lambda L\ll 1$, which is indeed the interesting
situation (since, typically, $L\sim 1/\Lambda$).

\subsection{From single scattering to the saturation momentum}
\label{sect-Qsat}

To further compute the DIS cross--section according to \eqnum{strucfunc}
we need to specify the metric perturbation $\hh(z,\bp)$ which enters the
scattering amplitude \eqref{T}. The crucial point for what follows is
that, as manifest on \eqnum{strucfunc}, the DIS process is only sensitive
to relatively small values of $z$, such that $z \lesssim 1/Q \ll
1/\Lambda$. In view of this and of the discussion in Sect.~\ref{sect-SW},
it is clear that, for this purpose, it is enough to retain the dominant
behavior of $\hh(z,\bp)$ near the boundary ($z\to 0$), which for any
acceptable shockwave metric scales like $z^4T_{--}$, cf. \eqnum{HREN}.
Thus our formalism makes it clear that the DIS process at strong coupling
involves the scattering off the nuclear energy density $T_{--}$. On the
supergravity side, this scattering involves multiple graviton exchanges
(here resummed in the eikonal approximation), with each factor of $z^4$
corresponding to one exchanged graviton. Hence, from now on we shall
simply take
 \beq\label{hTsimple}
  h(x^-,\bp,z) = \hh(\bp,z) \delta(x^-) = 2 \pi^2
  \frac{T_{--}}{N_c^2}\,z^4\,,
 \eeq
which is the approximate solution to the Einstein equation for the metric
component $h$ so long as $z \Lambda \ll 1$. One can cover both type of
shockwaves described in Sect.~\ref{sect-SW} by taking
 \beq\label{TmmE}
  T_{--}(x^-,\bp) = \Lambda^2 E f(\bp \Lambda) \delta(x^-)\,
  \equiv\,\tilde T_{--}(\bp)\delta(x^-)\,,
 \eeq
where $E$ is the total energy of the nucleus: for the type--I shockwave
in Eqs.~\eqref{Tmmsimple}--\eqref{fprof}, this reads $E=\gamma L
\Lambda^2 N_c^2$, whereas for the type--II one in \eqnum{Tgubser} one has
$E=p^+ N_c^2$ and we identify $z_*=1/\Lambda$.

Let us now consider the single scattering approximation which amounts to
expanding $\mathcal{T}(z,\bp)$ in \eqnum{T} to lowest order. By also
making use of \eqnum{hTsimple} the amplitude reads
 \beq\label{Tsingle}
 \mathcal{T}^{(1)}(z,\bp) = q^- \pi^2 \frac{\tilde T_{--}(\bp)}{N_c^2}\,z^4,
 \eeq
which is purely real, reflecting the fact that the single graviton
exchange generates no imaginary part (one cannot cut through a single
graviton propagator). Note that the single graviton exchange is
proportional to the energy density in the projectile and the target, and
is of order zero in $N_c^2$, because the target contains $N_c^2$ degrees
of freedom per unit volume (otherwise it would be suppressed like
$1/N_c^2$). Via the UV/IR correspondence $z \Leftrightarrow r_\perp$,
\eqnum{Tsingle} implies that a small dipole fluctuation of the
projectile, so like a gluon--gluon pair, with transverse size $r_\perp$
interacts with the target with an amplitude $\mcal{T}(r_\perp)\sim
r_\perp^4$, which vanishes much faster at small $r_\perp$ than at weak
coupling\footnote{This would not be true for the shockwave metric
selected in Ref.~\cite{Beuf:2009mk}, and which is obtained by taking
$c_2=0$ and $c_1\ne 0$ in \eqnum{bsolution}. In that case, the DIS
cross--section would involve only the first, constant, term in the
small--$z$ expansion \eqref{expbnd}, which via the UV/IR correspondence
would imply that the scattering amplitude remains constant as $r_\perp\to
0$.}  (where one rather has $\mcal{T}\sim r_\perp^2$ \cite{CGCreviews}).
As we shall later argue, this difference is due to the fact that, at
strong coupling, there are no high--$Q^2$ partons in the target
wavefunction, that the dipole could scatter off.

In this single--scattering approximation the polarization function
simplifies to
 \beq\label{sfsingle}
 \Pi^{(1)}(x,Q^2) = \frac{Q^6 \Lambda N_c^2}{32 x L}
 \int \dif z\, \dif^2\bp\,
 z \rmK_2^2(Q z)\,
 \frac{q^- \tilde T_{--}(\bp)}{N_c^2}\,z^4\,,
 \eeq
which is purely real and hence describes only elastic scattering. The
integrals in \eqnum{sfsingle} can be exactly perform and they yield the
same final result for both types of metric (for the type--I metric, the
integral over $b_\perp$ is performed according to \eqnum{fprof}), which
reads
 \beq\label{PI1}
 \Pi^{(1)}(x,Q^2) = \,\frac{N_c^2\Lambda^2 Q^2}{10 \,x^2}\,,
 \eeq
This is suppressed by a power of $\Lambda^2/ Q^2$ with respect to the
corresponding vacuum result in \eqnum{Pi0} and thus it is recognized as a
leading--twist effect. The fact that this is purely real, \emph{i.e.} the
structure function $F(x,Q^2)$ vanishes in the leading--twist
approximation, means that there are no point--like constituents in the
nuclear wavefunction at strong coupling, in agreement with previous
results in \cite{Polchinski:2002jw,HIM1,HIM2}. The $1/x^2$ rise in
\eqnum{PI1}, which on the supergravity side is clearly associated with
the graviton exchange, can be also understood in the original gauge
theory, as we explain now: in the OPE of the current--current correlator,
valid at high $Q^2$, the leading--twist operators with spin $n$ should
produce contributions which behave like $1/x^n$. At strong coupling, one
expects all such operators to acquire large and negative anomalous
dimensions\cite{Gubser:2002tv,Kotikov:2004er,Brower:2006ea}, of
$\order{\lambda^{1/4}}$, with the exception of the spin 2
energy--momentum tensor operator, which is protected by symmetry. Hence,
the only leading--twist contribution which is expected to survive at
strong coupling is the one which behaves like $1/x^2$, in
agreement\footnote{Via simple dimensional arguments, one can convince
oneself that the contribution of $T_{\mu\nu}\approx
T_{--}\delta_{\mu-}\delta_{\nu-}$ to the OPE of $\Pi(x,Q^2)$ can be
written as the product of $Q^2/x^2$ times a quantity which represents the
average energy density per unit transverse area in the nuclear target.
This quantity is of order $N_c^2\Lambda^2$, so the overall contribution
has indeed the parametric form shown in \eqnum{PI1}.} with \eqnum{PI1}.

Although it does not contribute directly to the structure function, the
single scattering contribution in \eqnum{PI1} allows us to derive the
sum--rule \eqref{SR}. Specifically, let us introduce the variable
$\nu\equiv 1/x$ and extend the polarization function $\Pi(\nu,Q^2)$ to
complex values for the variable $\nu$. Then, this function is expected to
be analytic everywhere in the complex $\nu$ plane except for two branch
cuts along the real axis, at $\nu>1$ and $\nu<-1$ (the physical region
for DIS and, respectively, the process related to DIS by crossing
symmetry; see, e.g., \cite{Peskin}). Then, via contour integration in the
complex $\nu$ plane, one can relate the behavior of this function near
$\nu=0$, where the OPE applies, to integrals (`moments') of the structure
function $F={\rm Im}\,\Pi$ along the branch cuts. In the present context
at strong--coupling, there is only one leading--twist contribution to
$\Pi(\nu,Q^2)$ --- the single--scattering piece
$\Pi^{(1)}(\nu,Q^2)\propto \nu^2Q^2$, cf. \eqnum{PI1} ---, meaning that
there is only one moment of the structure function which survives in the
high $Q^2$ limit\footnote{It is understood that one has to multiply the
moment by the ratio $\Lambda^2/Q^2$ before taking the limit
$Q^2\to\infty$, so like in \eqnum{SR}.}. Clearly, this moment is
proportional to the contribution of the energy--momentum tensor (the only
protected leading--twist operator) to $\Pi$. This contribution is
isolated as follows:
 \beq\label{SRder}
 \frac{N_c^2\Lambda^2 Q^2}{10}\,
 \,=\,\oint \frac{\rmd \nu}{2\pi i} \frac{\Pi^{(1)}}{\nu^3}
\,\simeq \,2\int_1^\infty\frac{\rmd \nu}{2\pi i}\frac{2i \,{\rm
Im}\,\Pi}{\nu^3}\,=\,\frac{2}{\pi}\int_0^1 \rmd x\, x\, F(x,Q^2),
 \eeq
where the contour in the first integral is a small circle surrounding the
origin which is then distorted in the complex plane in such a way to wrap
around the two branch cuts which give equal contributions. The
approximate sign in the second equality refers to the leading--twist
approximation. As anticipated, this procedure reproduces \eqnum{SR}.

The dominant contribution to the DIS structure function at high $Q^2$ is
of higher--twist order and comes from the two graviton exchange. This is
easily estimated as
 \beq\label{sfdouble}
 F^{(2)}(x,Q^2) &\,
 = &\frac{\pi^2 Q^6 \Lambda N_c^2}{32  x L}
 \int \dif z\, \dif^2\bp\, z \rmK_2^2(Q z)\,
 \left[\frac{q^- \tilde T_{--}(\bp)}{N_c^2}\,z^4\right]^2\,= \kappa \,\frac{16\pi}{7}\, \frac{N_c^2\Lambda^5 L}{x^3}\,,
 \eeq
with $\kappa = \pi \Lambda^2 \int d^2 b_\perp f^2$; in particular,
$\kappa = 1/8$ for a type--I shockwave with the exponential profile $f =
(1/2\pi)\exp(-\bp \Lambda)$ and $\kappa = 4/5$ for the type--II
shockwave. Since generated by cutting in between two graviton exchanges,
this contribution to the structure function can be characterized as being
{\em diffractive}. Similarly, by expanding the eikonal phase in \eqnum{T}
to higher orders, one generates diffractive contributions to $F(x,Q^2)$
of successively higher twist order and which increase faster and faster
with decreasing $x$ (as a power $1/x^{n+1}$ for the term corresponding to
$n$ graviton exchange). If one tries to use any single such a diffractive
contribution within the sum--rule \eqref{SR}, the ensuing integral is
ill--defined at its lower limit $x\to 0$. This reflects the fact that the
expansion of the eikonal amplitude makes sense only so long as the
exponent is much smaller than one. Thus, clearly, one expects a change of
regime when this exponent, which is the same as the single scattering
amplitude, becomes of $\order{1}$.

To study this change of regime, let us use \eqnum{TmmE} to rewrite
 \beq\label{T1approx}
 \mathcal{T}^{(1)} = q^- \pi^2\, \frac{E \Lambda^2 f(\bp \Lambda)}{N_c^2}\, z^4
  \sim \frac{\Lambda^3 L f(\bp \Lambda)}{Q^2 x}.
 \eeq
where we have also used the fact that $z \sim 1/Q$ together with
\eqnum{Bj}. (This estimate applies to both types of shockwaves with the
understanding that $L\sim 1/\Lambda \sim z_*$ for the type--II metric.)
The condition that $\mathcal{T}^{(1)}$ be of $\order{1}$ can be solved
for $Q^2$ at fixed $x$ and $\bp$, thus defining the {\em saturation
momentum} :
\beq\label{Qsimpact}
 Q_s^2(x,\bp) = \frac{\pi^2\Lambda^3 L}{2x}\,f(\bp\Lambda),
 \eeq
where the factor $\pi^2/2$ has been introduced for later convenience.
Note that the transverse profile $f(\bp\Lambda)$ of the nuclear energy
density directly transmits to the saturation momentum. In particular, for
a shockwave which is homogeneous in the transverse impact parameter space
one has $Q_s^2 =\pi L\Lambda^3/2x$, in agreement with
Refs.~\cite{HIM1,Mueller:2008bt}, but in disagreement with
Ref.~\cite{Albacete:2008ze}, where a rather surprising result for $Q_s$
was reported, which becomes independent of $x$ at high energy. As it
should be clear from the above analysis, and also from the corresponding
ones in Refs.~\cite{HIM1,Mueller:2008bt}, the $1/x$ rise of $Q_s^2$ at
small $x$ reflects the respective behavior of the one-graviton--exchange
scattering amplitude, and thus it seems unavoidable within this
gauge/gravity duality context, where the high energy scattering always
amounts to graviton exchanges.

Note finally that $\mathcal{T}^{(1)}(z\sim 1/Q,\bp)\sim
Q_s^2(x,\bp)/Q^2$, which reaches its maximal value at the center of the
`nucleus' ($\bp=0$). This makes it clear that the `twist' expansion of
the eikonal phase is an expansion in powers of $Q_s^2(x,0)/Q^2$ and is
appropriate in the high--$Q^2$ regime at $Q^2\gg Q_s^2(x,0)$. The
opposite regime, at $Q^2\lesssim Q_s^2(x,0)$, or equivalently $x \leq
x_s(Q)$ where,
 \beq\label{xs}
 x_s(Q)\, =\, \frac{\pi^2\Lambda^3 L}{2Q^2}f(0),
 \eeq
will be studied in the next subsection.

\subsection{Structure function at saturation}
\label{sect-sf}

Having identified the saturation momentum which marks the borderline
between weak and strong scattering, we shall now compute the structure
function in the `saturation region' at $Q^2\lesssim Q_s^2(x,0)$. This
will allow us to understand how the energy--momentum sum rule \eqref{SR}
is satisfied and speculate about a possible partonic interpretation.

Let us start by rewriting the scattering amplitude defined in
Eq.~\eqref{T} in terms of the saturation momentum introduced in the
previous subsection. It reads
 \beq\label{TQs}
 \mcal{T}(z,x,\bp) = \rmi \big\{1 -
 \exp \left[\rmi Q^2 Q_s^2(x,\bp) z^4\right] \big\}
 \eeq
In this high energy, or relatively low $Q^2$ regime, that we are
interested in, there is a central region of the nucleus which looks
`black' to the projectile. That is, for any given $z$ there is a `black
disk' radius $R_{\rm b}(z,Q,x)$ such that, for all $\bp \lesssim R_{\rm
b}$ the eikonal phase in \eqnum{TQs} can be neglected since rapidly
oscillating and the scattering amplitude becomes purely imaginary with a
magnitude equal to one, meaning that the dilaton component at radial
distance $z$ is completely absorbed in the shockwave. Recalling also
Eq.~\eqref{Qsimpact}, we see that this black disk radius is determined by
 \beq\label{bddef}
 \frac{f(R_{\rm b} \Lambda)}{f(0)} = \frac{1}{Q^2 Q_s^2(x,0) z^4}.
 \eeq
For given $x$ and $Q^2$, a black disk exists at the center of the
`nucleus' ($\bp=0$) only for those components of the dilaton which have
penetrated far enough in the radial direction, namely up to $z\ge z_0$
with $z_0 = 1/\sqrt{Q Q_s(x,0)}$. Physically, this means that the
transverse size $r_\perp\sim z$ of the partonic fluctuations of the
projectile should be large enough for their color dipolar interactions in
the target to be strong. Since on the other hand $z$ is restricted to
$z\lesssim 1/Q$ by the dilaton wavefunction, it is clear that a black
disk can form only when the energy is high enough for the condition
$Q^2\lesssim Q_s^2(x,0)$ to be satisfied. Once that the black disk
appears at the center of the `nucleus' (for a given $z\ge z_0$) it
rapidly occupies the whole central region of the shockwave, which is
essentially flat in $\bp$. When further increasing the energy, the black
disk extends al larger values of $\bp$, at a rate which depends upon the
transverse profile of the shockwave.

For example, for a type--II shockwave the black disk radius will be given
by
 \beq\label{bdtype2}
 (R_{\rm b} \Lambda)^2 = [Q^2 Q_s^2(x,0) z^4]^{1/3} - 1\,.
 \eeq
This exhibits a rapid increase with the energy, proportional to
$(1/x)^{1/3}$, which reflects the power law tail in $\bp$ of the
respective energy density profile. On the contrary, a type--I shockwave
with an exponential tail at large $\bp$ will have a black disk radius
which at high energy increases with $1/x$ only logarithmically, that is
 \beq\label{bdtype1}
 (R_{\rm b} \Lambda)^2 = \ln^2[Q^2 Q_s^2(x,0) z^4].
 \eeq
For high enough energy, the total cross section \eqref{sigma} for a given
value of $z$ will be proportional to the area of the respective black
disk. (Indeed, outside the black disk, at $\bp \gg R_{\rm b}(z,Q,x)$, the
amplitude is very small and rapidly decreasing with $\bp$.) So, clearly,
the cross--section associated to the type--II shockwave will rise like a
power of the energy, in violation of the Froissart bound, whereas that
corresponding to a type--I shockwave with exponential tail will saturate
the Froissart bound at high enough energy. In that sense, the type--II
shockwave is a more realistic model for a nucleus in QCD. In fact, the
above mechanism for the emergence of the Froissart bound is exactly the same
as expected in QCD: the competition between the power--law increase of
the scattering amplitude with the energy at a given $\bp$ and the
exponential decrease of the nuclear matter distribution at large $\bp$.

The previous considerations lead to the following expression for the
structure function at $Q^2< Q_s^2(x,0)$ :
 \beq\label{Fz}
 F(x,Q^2) = \frac{N_c^2}{16 \pi}\,\frac{Q^6 }{x \Lambda L}
 \int_{z_0}^{\infty} \dif z \, z\, \rmK_2^2(Q z)
 \,[R_{\rm b}(z,Q,x) \Lambda]^2.
 \eeq
The Bessel function will effectively cut off all contributions coming
from $z \gtrsim 1/Q$, so that the integration in Eq.~\eqref{Fz} is in
practice supported only in the interval
 \beq\label{zinterval}
 \frac{1}{\sqrt{Q Q_s(x,0)}} \,\lesssim\, z \, \lesssim\, \frac{1}{Q}\,.
 \eeq
However, the dilaton wavefunction diverges strongly at small $z$, more
precisely $\rmK_2^2(Q z) \simeq 4/Q^4 z^4$. Thus for any profile
$f(\bp\Lambda)$ which leads to a black disk area $(R_{\rm b} \Lambda)^2$
larger than $z^2$ at $z\ll 1$, therefore including the two cases in
Eqs.~\eqref{bdtype2} and \eqref{bdtype1}, the integration is dominated by
its lower limit. This comes as a surprise, since one might have expected
to dominant contribution to come from $z \sim 1/Q$, as happens in the
case of the $\mcal{R}$--current (see Sect.~\ref{sect-R} below), and also
in perturbative QCD (after identifying $z$ with the size of the dipole
fluctuation of the virtual photon). Coming back to the case under study,
we see that for $z \sim z_0$ the black disk area $(R_{\rm b}\Lambda)^2$
is a number of $\order{1}$ and thus the integration is simply
proportional to $1/Q^4 z_0^2 = Q_s(x,0)/Q^3$. So, quite remarkably, the
structure function for the dilaton DIS is {\em not} sensitive to the
expansion of the black disk with increasing energy --- rather, it is
controlled by the central part of the black disk at $\bp\lesssim
1/\Lambda$ ---, and hence it is not affected, e.g., by the violation of
the Froissart bound in the case of the type--II shockwave. This should be
contrasted with the corresponding situation for the $\mcal{R}$--current,
to be described in Sect.~\ref{sect-R}.

As we show in Appendix \ref{app-pre}, deeply at saturation ($Q^2 \ll
Q_s^2(x,0)$), we can also determine the precise normalization of the
structure function, and we finally have
 \beq\label{sfdeep}
 F(x,Q^2) = \kappa\, \frac{N_c^2}{16 \sqrt{2\pi}}\,\frac{Q^4 }{x \Lambda L}\,
 \frac{Q_s(x,0)}{Q},
 \eeq
with $\kappa$ a number which depends only on the profile of the shockwave
under consideration.

It is straightforward to check that in the transition region at $Q \sim
Q_s(x)$, or equivalently $x \sim x_s(Q)$, the above result in
Eq.~\eqref{sfdeep} is parametrically consistent with the respective
extrapolation of the dominant result at large--$x$, \emph{i.e.}~the
2--graviton exchange contribution in \eqnum{sfdouble}:
 \beq\label{Fxs} x_s F_s(x_s,Q^2)\,\sim\,
 \frac{Q^4 N_c^2}{\Lambda L} \,\sim\,x_s F^{(2)}(x_s,Q^2)\,.\eeq
Of course, for $x \sim x_s(Q)$ all the terms in the twist--expansion
become parametrically of the same order and our various approximations
fail to apply there, except for parametric estimates.

By using these estimates, one can now verify that the energy--momentum
sum rule \eqref{SR} is indeed satisfied parametrically. The integral over
$x$ in the l.h.s. of \eqnum{SR} is dominated by $x\sim x_s(Q)$, since for
$x\gg x_s$ the 2--graviton exchange contribution in \eqnum{sfdouble} is
rapidly decreasing with $x$, whereas for $x\lesssim x_s$, we have $x
F(x,Q^2)\sim 1/\sqrt{x}$. Thus, by making use of Eqs.~\eqref{Fxs} and
\eqref{xs} one can write
 \beq
 \int_0^1 \rmd x\, x\, F(x,Q^2)\,\sim\, x_s^2\, F(x_s,Q^2)\,\sim
 \,\frac{Q^4 N_c^2x_s}{\Lambda L} \,\sim\,Q^2
 N_c^2 \Lambda^2.
\eeq

Let us conclude this discussion by suggesting a possible partonic
interpretation for the previous results. The standard OPE analysis for
the correlator \eqref{JJDIS} shows that, in the weak coupling regime and
for sufficiently high $Q^2$, the quantity $xF(x,Q^2)/Q^2$ has the meaning
of the gluon distribution in the target (since gluons are the partons
which directly couple to the relevant operator $J=(1/4)F^2$). Namely, it
is proportional with the number of gluons per unit transverse area having
longitudinal momentum fraction $x$ and transverse momenta
$k_\perp\lesssim Q$. Then what about the previous results at strong
coupling ? Since the energy--momentum sum--rule is controlled by $x\sim
x_s(Q)$, it is natural to look at \eqnum{Fxs}, which implies $x_s
F_s(x_s,Q^2)/Q^2 \sim {Q^2 N_c^2}$. As we shall argue in more detail in
Sect.~\ref{sec-F2}, where a similar result occurs in the context of the
$\mathcal{R}$--current, this estimate is indeed consistent with a
partonic picture, in which the partons (here, gluons) are distributed
along the saturation line with occupation numbers of order one
\footnote{At this stage, it might be useful to remember that, in the
saturation region at $x\le x_s(Q)$, the partonic interpretation of the
DIS structure functions does not rely on the validity of the twist
expansion (and the dominance of the twist--two operators) not even at
weak coupling \cite{CGCreviews}.}. One the other hand, there are clearly
no partons at $x\gg x_s(Q)$, as obvious from the lack of leading--twist
contributions to $F(x,Q^2)$ in that region.

\section{DIS off the shockwave: the $\mathcal{R}$--current}
\label{sect-R}

Let us now turn to the problem of deep inelastic scattering of an
$\mcal{R}$-current $J_{\mu}$ off the shockwave. As discussed in the
beginning of Sect.~\ref{sect-D} this is a conserved current in
$\mcal{N}=4$ SYM which is carried by the fermionic and scalar fields. It
is a bilinear in both of these types of fields, whose precise form is not
necessary for our discussion (it is given, for example, in \cite{HIM2}),
and it has mass dimension 3.

\subsection{General equations: from the 5D action to the polarization tensor}

We will be interested in calculating the polarization tensor
 \beq\label{Rmunu}
  \Pi^{\mu\nu}(q) = \frac{\gamma \Lambda^2}{L} \int \dif^4 r\,
  \dif b^- \,\dif^2\bp\, \rme^{\rmi q \cdot r} \ \rmi \Theta(x_0)
   \lan[J^{\mu}(b+r/2),J^{\nu}(b-r/2)] \ran,
 \eeq
which is the analogous to Eq.~\eqref{PI} and where the notation is the
same as in the previous section. The bulk field dual to the
$\mcal{R}$-current $J_{\mu}$ is the SO(6) gauge field $A^{m}$, where we
have suppressed the fixed color index of this vector field corresponding
to the fixed index of the $\mcal{R}$-current. To compute $\Pi^{\mu\nu}$
(see \eqnum{PImunu} below), we need only the quadratic part of the
respective Yang--Mills action in the AdS$_5$ background. It reads
 \beq\label{Smax}
 S = -\frac{N_c^2}{64 \pi^2 R} \int \dif^4 x\,\dif z\,
 \sqrt{-g}\, g^{mp} g^{nq} F_{mn} F_{pq},
 \eeq
with $F_{mn} = \del_m A_n - \del_n A_m$. According to the AdS/CFT
correspondence the 2--point function of the current appearing in the
integrand Eq.~\eqref{Rmunu} is equal to
 \beq\label{PImunu}
  \Pi^{\mu\nu}(x,y) = \frac{\delta^2 S_{\rm cl}}{\delta\mcal{A}_{\mu}(x)
  \delta\mcal{A}_{\nu}(y)}\bigg|_{\mcal{A}_{\mu}=0}.
 \eeq
In the above, $S_{\rm cl}$ is the action evaluated on the solution to the
equations of motion
 \beq\label{Maxwell}
 \del_{m}(\sqrt{-g}\,g^{mp} g^{nq} F_{pq}) = 0,
 \eeq
\emph{i.e.} the Maxwell equations in curved space--time, with the
following boundary conditions
 \beq
  \lim_{z \to 0} A_{\mu}(z,x) = \mcal{A}_{\mu}(x)
  \quad \mathrm{and} \quad \lim_{z \to 0} A_{z}(z,x) =0.
 \eeq
It will be convenient to adopt the gauge condition $A_z = 0$.
Furthermore, as in the case of the dilaton, we shall assume that the
boundary field is a plane wave of the form
 \beq
  \mcal{A}_{\mu}(x) = \rme^{-\rmi (q^-x^+ + q^+ x^-)} \tilde{\mcal{A}}_{\mu},
 \eeq
so that the components of $\tilde{\mcal{A}}_{\mu}$ are pure numbers. Now
let us integrate Eq.~\eqref{Smax} by parts. Using the Maxwell equations
\eqref{Maxwell} only the boundary term survives to give
 \beq
 S_{\rm cl} = -\frac{N_c^2}{32 \pi^2 R} \int \dif^4 x\, \dif z\,
 \del_m(\sqrt{-g}\, g^{mp} g^{nq} F_{pq} A_n).
 \eeq
Substitution of our shockwave metric into the above leads to
 \beq\label{Smaxcl}
 S_{\rm cl} = \frac{N_c^2}{32 \pi^2} \int \dif^4 x\, \frac{1}{z}\,
 \left(- A_+ A_{-}^{\prime} - A_- A_{+}^{\prime} + A_i A_{i}^{\prime}\right)\Big|_{z=0},
 \eeq
where a prime represents differentiation with respect to $z$, the index
$i=1,2$ refers to the transverse components and we have dropped a term
$-h A_+ A_{+}^{\prime}$ inside the parenthesis which does not contribute
since $h \propto z^4$ for small $z$.

For simplicity, from now on we will restrict to the case where the fields
are independent of $x_{\perp}$, which is the situation when the shockwave
is homogeneous in transverse directions: $\del_i h = 0$. Furthermore,
since the shockwave metric in Eq.~\eqref{metric} does not depend on
$x^+$, we can restrict ourselves to a single Fourier mode of the gauge
field as follows
 \beq
 A_{\mu}(z,x^+,x^-) = \rme^{-\rmi q^- x^+} A_{\mu}(z,x^-).
 \eeq
The Maxwell equations for $n = -,z,i$ become
 \beq\label{mMaxwellh}
 &&(z \del_z z^{-1} \del_z + \rmi q^{-} \del_-) A_+
 = (q^{-})^2 A_-,
 \\*[0.3cm] \label{zMaxwellh}
 &&(\del_- - \rmi q^- h) A_+^{\prime} -= \rmi q^- A_-^{\prime},
 \\*[0.3cm] \label{iMaxwellh}
 &&(z \del_z z^{-1} \del_z + 2 \rmi q^{-} \del_-)A_i = - (q^{-})^2 h A_i.
 \eeq
Notice that in the vacuum case $h=0$, the $x^-$ dependence of the field
is that of a plane wave, that is $\rme^{-\rmi q^+ x^-}$.  The first two
equations are coupled. In order to solve them  we differentiate
\eqref{mMaxwellh} with respect to $z$ and we make use of
\eqref{zMaxwellh} to arrive at the differential equation which determines
$A_+^{\prime}$ :
\beq\label{AplusEOMh}
 (\del_z z \del_z z^{-1} + 2 \rmi q^{-} \del_-)
 A_+^{\prime} = - (q^{-})^2 h A_+^{\prime}.
 \eeq
As in the dilaton case, we separate the total field into a vacuum piece
and a scattering piece
 \beq
 A_{\mu}(z,x^-) = A_{\mu}^{(0)}(z,x^-) + A_{\mu}^{(s)}(z,x^-),
 \eeq
with $A_{\mu}^{(0)}(z,x^-)$ satisfying the vacuum version of the Maxwell
equations which are obtained by setting $h=0$. We shall construct
$A_{\mu}(z,x^-)$ so that its boundary condition at $z=0$ is fully encoded
in the vacuum field, that is
  \beq
  \lim_{z \to 0} A_{\mu}^{(0)}(z,x^-) = \mcal{A}_{\mu}(x^-)
  \quad \mathrm{and} \quad\lim_{z \to 0} A_{\mu}^{(s)}(z,x^-) = 0.
  \eeq
Now we define the boundary to bulk propagator $D^{\mu\nu}$ according to
  \beq
  A_{\mu}^{(s)}(z,x^-) = \int \dif y^-\, D_{\mu}{}^{\nu}(z,x^-,y^-)
   \mcal{A}_{\nu}(y^-),
  \eeq
which must be determined by solving the equations of motion. Then the
contribution of the scattering piece to the classical action
\eqnum{Smaxcl} is obtained from $-A_+^{(0)} A_-'^{(s)} - A_-^{(0)}
A_+'^{(s)} + A_i^{(0)} A_i'^{(s)}$ and reads
  \beq
  S_{\rm cl} - S_0 = V\,\frac{N_c^2}{32 \pi^2} \int \dif x^- \dif y^- \mcal{A}_{\mu}(x^-) \mcal{A}_{\nu}(y^-)\frac{1}{z}\, \del_z D^{\mu\nu}(z,x^-,y^-) \bigg |_{z=0},
  \eeq
with the volume factor $V$ arising from the integration over $x^+$ and
$x_{\perp}$. Functional differentiation according to Eq.~\eqref{PImunu}
and use of the same prescription as the one used in the dilaton case
(cf.~the discussion just after Eq.~\eqref{2D}) leads to\footnote{Notice
that there is a factor $1/V$ for each functional differentiation since
$\delta/\delta\mcal{A}(x) = (1/V) \delta/\delta\mcal{A}(x^-)$.}
 \beq
 \Pi_{\rm R}^{\mu\nu} = \frac{1}{V}\,\frac{N_c^2}{16 \pi^2}\,
 \frac{1}{z}\,\del_z D^{\mu\nu}(z,x^-,y^-) \bigg |_{z=0}.
 \eeq
Substituting into Eq.~\eqref{Rmunu} we finally arrive at
 \beq\label{Rmunufinal}
 \Pi^{\mu\nu}(q) = \frac{\gamma}{L}\,\frac{N_c^2}{16 \pi^2}
 \int \dif x^- \dif y^-\, \rme^{\rmi q^+(x^- - y^-)}\,\frac{1}{z}\,
 \del_z D^{\mu\nu}(z,x^-,y^-) \bigg |_{z=0}.
 \eeq
Let us notice here that not all the components of the propagator
$D^{\mu\nu}$, and therefore of the tensor $\Pi^{\mu\nu}$, are
independent. In fact, only two components are independent since
$\Pi^{\mu\nu}$ can be decomposed as
 \beq\label{Rmn}
 \Pi^{\mu\nu}(q) = \left( \eta^{\mu\nu} - \frac{q^{\mu} q^{\nu}}{Q^2} \right)
 \Pi_1(x,Q^2) +
 \left(p^{\mu} - \frac{p \cdot q}{Q^2}\, q^{\mu}\right)
 \left(p^{\nu} - \frac{p \cdot q}{Q^2}\, q^{\nu}\right) \Pi_2(x,Q^2),
 \eeq
which follows from general symmetries (namely the current conservation
together with the time--reversal symmetry $\Pi^{\mu\nu}(q)
=\Pi^{\mu\nu}(-q)$) and can be explicitly verified by calculating one by
one all the components of the propagator $D^{\mu\nu}$. Hence we will just
need to determine $\Pi^{ii}$ and, for example, $\Pi^{--}$, since the
remaining non--vanishing components will be then determined as
 \beq\label{R++R+-}
 \Pi^{++} = \frac{Q^4}{4 (q^{-})^4}\, \Pi^{--} \quad \mathrm{and} \quad
 \Pi^{+-} = \Pi^{-+} = \frac{Q^2}{2 (q^{-})^2}\, \Pi^{--}.
 \eeq

\subsection{The classical solution}

The solution to the equations of motion is constructed in the same way as
in the dilaton case. Resumming the graviton exchanges to all orders we
find for the scattering part of the transverse field
\beq\label{Ais}
 A_i^{(s)}(x^-,z) =
 -2 q^-\int
 \frac{\dif \acute{z}}{\acute{z}}\,G_{\rm T}
 (z,\acute{z}; x^-)\,
 \mcal{T}(\acute{z})
 A_{i}^{(0)}(0, \acute{z}),
 \eeq
where the scattering amplitude $\mcal{T}(\acute{z})$ has already been
defined in Eq.~\eqref{T} and with the Green's function $G_{\rm T}$
satisfying in momentum space
 \beq\label{GT1}
 (z \del_z z^{-1} \del_z - K^2) G_{\rm T}(z,\acute{z};K^2) = z \delta(z-\acute{z}),
 \eeq
with the notation $K^2 = -2 k^+ q^-$. The precise form of the Green's
function is given in Appendix \ref{app-G}. Now we can see that the
respective boundary to bulk propagator reads
 \beq\label{Dii}
 D^{ii}(z,x^-,y^-) = - 2 q^- \int \frac{\dif\acute{z}}{\acute{z}}
 \,G_{\rm T} (z,\acute{z}; x^-) \mcal{T}(\acute{z})  D_0^{ii}(\acute{z},-y^-),
 \eeq
where $D_0^{ii}$ is the corresponding boundary to bulk propagator in the
vacuum and in momentum space is given by
 \beq\label{Dii0}
 D^{ii}_0(z,K) = K z \rmK_1(Kz).
 \eeq
Substituting the above into Eq.~\eqref{Rmunufinal} we can immediately
perform the integrations over the longitudinal coordinates which will
select the $q^+$ component of both the Green's function $G_{\rm T}$ and
the propagator $D^{ii}_0$. We then have (recall that $Q^2=-2q^+q^->0$)
  \beq
  \Pi^{ii} = -\frac{N_c^2}{16 \pi^2}\, \frac{Q^2}{x\Lambda L} \int
  \frac{\dif \acute{z}}{\acute{z}}\,
  \frac{1}{z}\, \del_z G_{\rm T}(z,\acute{z};Q^2) \Big |_{z=0}
  \mcal{T}(\acute{z}) D^{ii}_0(\acute{z},Q).
  \eeq
Finally by making use of
 \beq
 -\frac{1}{z}\,\del_z G_{\rm T}(z,\acute{z};Q^2) \Big |_{z=0}  = Q\acute{z} \,
 \rmK_1(Q \acute{z}),
 \eeq
and Eq.~\eqref{Dii0}, and changing the name of the integration variable
from $\acute{z}$ to $z$, we arrive at
 \beq\label{Rii}
 \Pi^{ii} = \frac{N_c^2}{16 \pi^2}\, \frac{Q^4}{x\Lambda L} \int
 \dif z\, z\, \rmK_1^2(Q z)\, \mcal{T}(z).
 \eeq
We recognize here a structure similar to that in \eqnum{strucfunc}: once
again, the Bessel function $\rmK_1^2(Q z)$ plays the role of the
wavefunction squared for the partonic fluctuation of the
$\mcal{R}$--current (namely, a fluctuation with transverse size $\Delta
x_\perp\sim z$). Interestingly, it is the same Bessel function which
determines the wavefunction of the quark--antiquark fluctuation of a
space--like photon in lowest--order perturbative QCD. The same remark
applies to the structure of $\Pi^{--}$, that we now compute.

Namely, the scattering part of the $A_+$ component of the gauge field is
found as
 \beq\label{Apluss}
 A_+^{\prime(s)}(x^-,z) =
 -2 q^-\int
 \frac{\dif \acute{z}}{\acute{z}}\,G_{\rm L}
 (z,\acute{z}; x^-)\,
 \mcal{T}(\acute{z})
 A_{+}^{\prime(0)}(0, \acute{z}),
 \eeq
with the Green's function $G_{\rm L}$ satisfying in momentum space
 \beq\label{GL1}
 (\del_z z \del_z z^{-1} - K^2) G_{\rm L}(z,\acute{z};K^2) = z \delta(z-\acute{z}),
 \eeq
and again the precise form of the Green's function is given in Appendix
\ref{app-G}. Now we need to be careful when we express the vacuum field
$A_{+}^{\prime(0)}(0, \acute{z})$ in terms of the boundary values. We
have
 \beq
 A_{+}^{\prime(0)}(x^-, z) = \int \dif y^-\, \del_z D_0^{-\mu}(z,x^-,y^-)
  \mcal{A}_{\mu}(y^-),
 \eeq
where the upper index $\mu$ can be either $+$ or $-$. Both components of
the boundary to bulk propagator $\del_z D_0^{-\mu}(z,x^-,y^-)$ satisfy
the vacuum version of Eq.~\eqref{AplusEOMh} and in momentum space they
are given by
 \beq
 \del_z D_0^{-\mu}(z,K) = \frac{1}{2}\,K^2 z\, \rmK_0(K z) c^{\mu}
 \quad \mathrm{with} \quad c^+ = 1 \quad \mathrm{and} \quad c^- = \frac{2 (q^-)^2}{Q^2}.
 \eeq
The coefficients $c^{\mu}$ have been determined by plugging the above
equations into the left hand side of Eq.~\eqref{mMaxwellh} and then
evaluating both sides of that equation on the boundary. Since we would
like to calculate $\Pi^{--}$ we shall need here only the $D_0^{--}$
component, but it is already obvious from the above equation why the
ratio $\Pi^{+-}/\Pi^{--}$ is given by the second equation in
Eq.~\eqref{R++R+-}. Now we are ready to give the boundary to bulk
propagator which reads
 \beq\label{D--}
 \del_z D^{--}(z,x^-,y^-) = - 2 q^- \int \frac{\dif\acute{z}}{\acute{z}}\,
 G_{\rm L} (z,\acute{z}; x^-) \mcal{T}(\acute{z})  \del_{\acute{z}}
 D_0^{--}(\acute{z},-y^-).
 \eeq
We proceed as before and we plug into Eq.~\eqref{Rmunufinal} to obtain
  \beq
  \Pi^{--} = -\frac{N_c^2}{16 \pi^2}\, \frac{Q^2}{x\Lambda L} \int
  \frac{\dif \acute{z}}{\acute{z}}\, \frac{1}{z}\,  G_{\rm L}(z,\acute{z};Q^2)
  \Big |_{z=0}
  \mcal{T}(\acute{z}) \del_{\acute{z}} D^{ii}_0(\acute{z},Q).
  \eeq
Using $-(1/z)G_{\rm L}(z,\acute{z};Q^2) |_{z=0} = \acute{z}\,
\rmK_0(Q\acute{z})$ and then letting $\acute{z} \to z$ we finally obtain
 \beq\label{R--}
 \frac{Q^2}{(q^-)^2}\,\Pi^{--} =
 \frac{N_c^2}{16 \pi^2}\, \frac{Q^4}{x\Lambda L} \int \dif z\,
 z\, \rmK_0^2(Q z) \mcal{T}(z).
 \eeq

\bigskip
\subsection{Structure functions and their partonic interpretation}
\label{sec-F2}

We are finally in a position to calculate the Lorentz invariant structure
functions $F_1$ and $F_2$. In general they are given by
 \beq\label{FR}
 F_1 = \frac{1}{2\pi}\, {\rm Im}\, \Pi_1,
 \qquad
 F_2 = -\frac{p\cdot q}{2\pi}\, {\rm Im} \,\Pi_2.
 \eeq
Focusing on the ($--$) and ($ii$) components in \eqref{Rmn} we can
express $\Pi_1$ and $\Pi_2$, and therefore the structure functions, in
terms of the components $\Pi^{--}$ and $\Pi^{ii}$ that we have already
calculated. It is also customary and convenient to introduce the
transverse and longitudinal structure functions $F_{\rm T}$ and $F_{\rm
L}$. Let us summarize here our results for all these various structure
functions:
 \beq\label{FT}
 F_{\rm T} &=& 2 x F_1 = \frac{x}{\pi}\,{\rm Im} \Pi^{ii} =
 \frac{N_c^2}{16 \pi^3}\, \frac{Q^4}{\Lambda L} \int
 \dif z\, z\, \rmK_1^2(Q z)\, \mcal{T}(z),
 \\ \label{FL}
 F_{\rm L} &=& \frac{x}{\pi}\,{\rm Im} \frac{Q^2}{(q^-)^2}\,\Pi^{--} =
 \frac{N_c^2}{16 \pi^3}\, \frac{Q^4}{\Lambda L} \int \dif z\,
 z\, \rmK_0^2(Q z) \mcal{T}(z),
 \\ \label{F2}
 F_2 &=&  F_{\rm T} + F_{\rm L}.
 \eeq
These results are in agreement with those in Ref.~\cite{Mueller:2008bt};
there is in fact a mismatch by an overall factor $\Lambda L$, but this is
presumably related to the use of different conventions in the
normalization of the current--current correlator.

We have already seen that the scattering amplitude $\mcal{T}$ that
emerged during the calculation is the same as the one appeared in the
dilaton case. Thus the saturation scale will also be the same in the case
we are currently considering. This is natural since the saturation scale
is a property of the nuclear target and hence it should be the same
whether we probe it with the gluon current or with the
$\mcal{R}$--current. Recalling that $Q_s^2 = \pi  \Lambda^3 L/2 x$ (in a
homogeneous situation) we can write the amplitude $\mcal{T}$ as
 \beq\label{TmQs}
 \mcal{T}(z,Q) = \rmi \left[1 - \exp\left(\rmi\,Q^2 Q_s^2\, z^4
 \right)\right].
 \eeq

Since it not possible to obtain an analytic expression for the structure
functions for arbitrary value of $x$ and $Q^2$, we shall separately
consider the two limiting cases.

({\tt i}) For $Q \gg Q_s(x)$ (or, equivalently, $x\gg x_s(Q)$, with
$x_s(Q)$ defined in \eqnum{xs}) we need to expand the exponential in
\eqref{TmQs} to second order since the first order term will not
contribute to the imaginary part of the tensor $\Pi^{\mu\nu}$. The upper
limit in the $z$--integration is $1/\sqrt{Q Q_s} \gg 1/Q$, so it can be
set equal to $\infty$. We find
 \beq\label{Fhigh}
 F_{\rm T}= \frac{16 N_c^2}{7\pi}\frac{\Lambda^2}{x}\,\frac{Q_s^2}{Q^2}, \qquad
 F_{\rm L}=  \frac{64 N_c^2}{35\pi}\,\frac{\Lambda^2}{x}\,\frac{Q_s^2}{Q^2}
 \quad \mathrm{and} \quad
 F_2 = \frac{144 N_c^2}{35\pi}\frac{\Lambda^2}{x}\,\frac{Q_s^2}{Q^2}.
 \eeq
As in the dilaton case, these contributions are of higher--twist order
(in the context of the $\mcal{R}$--current, a leading--twist contribution
would be independent of $Q^2$ at large $Q^2$, up to logarithms), which
reflects the absence of point--like constituents in the nucleus.

({\tt ii}) For $Q \ll Q_s(x)$ (or, equivalently, $x\ll x_s(Q)$), we set
the exponential in \eqref{TmQs} equal to zero and we restrict the
integration to $z \gtrsim 1/\sqrt{Q Q_s}$. In the longitudinal sector,
and since $\rmK_0$ diverges only logarithmically at small $z$, the
integration is dominated by the region $z \sim 1/Q$, so that the lower
limit can be safely set equal to zero. This is not the case in the
transverse sector where the integration is sensitive to small values.
Since $\rmK_1(Q z) \simeq 1/Q z$ for small $z$, there is a
logarithmically enhanced contribution when compared to $F_{\rm L}$.
Therefore the transverse sector dominates in $F_2$, albeit this dominance
is only logarithmic. We get
 \beq\label{Flow}
 F_2 \simeq F_{\rm T} = \frac{N_c^2}{64\pi^3}\,\frac{Q^2}{\Lambda L}\,
 \ln\frac{Q_s^2}{Q^2}
 \quad \mathrm{and} \quad F_{\rm L} =  \frac{N_c^2}{32\pi^3}\,\frac{Q^2}
 {\Lambda L}.
 \eeq
Remarkably, the structure functions at low $Q^2$ have the same parametric
form as for a proton in QCD at weak coupling, except for the replacement
of the baryonic color factor $N_c$ corresponding to a proton by a factor
$N_c^2$ corresponding to the density of degrees of freedom in the
shockwave.

Even though the above expressions in Eqs.~\eqref{Fhigh} and \eqref{Flow}
are strictly valid for $Q \gg Q_s$ and $Q \ll Q_s$ respectively, they
become of the same order when extrapolated to $Q \sim Q_s$ and therefore
they can be parametrically trusted even for $Q$ around $Q_s$. This also
means that for $Q \sim Q_s$ the transverse and longitudinal structure
functions are of the same order.

So far, our analysis in the $\mcal{R}$-current case has been done for a
homogeneous, in the transverse space, shockwave. However, given the close
analogy of the structure functions in Eqs.~\eqref{FT} and \eqref{FL} to
the dilaton structure function in Eq.~\eqref{strucfunc} and the common
form of the scattering amplitude in the two cases, it is not unreasonable
to extend the discussion of Sect.~\ref{sect-sf} to the problem of an
inhomogeneous shockwave probed by an $\mcal{R}$--current. In such a
situation, and according to Eq.~\eqref{Fz}, we would write the transverse
and longitudinal structure functions at saturation ($Q^2 \ll Q_s^2(x,0)$)
as
 \beq\label{FTLz}
 F_{\rm T,L}(x,Q^2) = \frac{N_c^2}{8 \pi^2}\,\frac{Q^4}{\Lambda L}
 \int_{z_0}^{\infty} \dif z \, z\, \rmK_{1,0}^2(Q z)
 \,[R_{\rm b}(z,Q,x) \Lambda]^2,
 \eeq
with $R_{\rm b}$ and $z_0$ as in Sect.~\ref{sect-sf}. But, in contrast to
the dilaton case, the $z$-integration is now dominated by $z \sim 1/Q$ in
the longitudinal sector, while it is only logarithmically sensitive to
the lower limit in the transverse sector (like in the homogeneous case),
a behavior similar to the one in QCD. For the exponential profile leading
to the black disk radius in Eq.~\eqref{bdtype1}, we find
 \beq
 F_2 \simeq F_{\rm T} = \frac{N_c^2}{192 \pi^2}\,\frac{Q^2}{\Lambda L}\,
 \ln^3 \frac{Q_s^2(x,0)}{Q^2}
 \quad \mathrm{and} \quad
 F_{\rm L} = \frac{N_c^2}{16 \pi^2}\,\frac{Q^2}{\Lambda L}\,
 \ln^2 \frac{Q_s^2(x,0)}{Q^2},
 \eeq
where we see again the extra logarithmic enhancement in the transverse
structure function.

In what follows, we shall argue that the results in \eqnum{Flow} have a
natural physical interpretation in terms of parton saturation in the
boundary gauge theory. Our discussion will closely follow that of the
plasma structure functions at strong coupling, as originally presented in
Ref.~\cite{HIM2}. To that purpose, two more ingredients turn out to be
extremely useful:

\bigskip

\texttt{(a)} The associated energy--momentum sum rules, which can be
derived via the same strategy as in the dilaton case (recall the
discussion of \eqnum{SRder}). Namely, one first computes the
leading--twist, or single scattering, contributions to the Lorentz
invariants $\Pi_1(x,Q^2)$ and $\Pi_2(x,Q^2)$, which are real quantities
as anticipated:
 \beq
 \Pi_1^{(1)} = \frac{N_c^2}{20}\,\frac{\Lambda^2}{x^2}
 \quad \mathrm{and} \quad
 \Pi_2^{(1)} = \frac{11 N_c^2}{15}\,\frac{\Lambda^2}{Q^2},
 \eeq
Then one uses analyticity and contour integration in the complex
$\nu\equiv 1/x$ plane to finally deduce
 \beq\label{sumrules}
 \int_{0}^{1} \dif x\, x F_1(x,Q^2) = \frac{N_c^2}{80}\,\Lambda^2
 \quad \mathrm{and} \quad
 \int_{0}^{1} \dif x\, F_2(x,Q^2) = \frac{11 N_c^2}{120}\,\Lambda^2\,,
 \eeq
which should more properly be understood as limiting values when
$Q^2\to\infty$ (since, as in the dilaton case, we have neglected the
higher--twist contributions to $\Pi(\nu,Q^2)$ near $\nu=0$). Examining
Eqs.~\eqref{Fhigh} and \eqref{Flow}, we can verify that these sum rules
are parametrically satisfied by our previous estimates for the structure
functions, and that the integrals are dominated by $x \sim x_s$, as
expected. In fact, by using the exact, integral, expressions for
$\Pi^{ii}$ and $\Pi^{--}$ derived in the previous subsection, one can
explicitly evaluate the above integrals of $F_1$ and $F_2$ and thus check
not only their high--$Q^2$ limit shown in \eqnum{sumrules}, but also the
subleading corrections at large $Q^2$. This will be detailed in Appendix
\ref{app-sr}.

Once again, the sum--rules \eqref{sumrules} extract the contribution of
the energy--momentum tensor to the operator product expansion for the
current--current correlator. In view of this and of the mass dimensions
of the above integrals, it is natural to interpret them as the energy per
unit length in the nucleus rest frame.

\bigskip
\texttt{(b)} The expression of the gluon distribution at saturation in
perturbative QCD, which reads, parametrically \cite{CGCreviews}
 \beq\label{xGx}
 \frac{xG(x,Q^2)}{\pi R^2}\,\sim\,\frac{1}{\lambda}\,N_c^2\,Q^2\,
 \ln\frac{Q_s^2(x)}{Q^2}\,,\eeq
where we now assume $\lambda\equiv g^2N_c\ll 1$ and $N_c\gg 1$. (The last
assumption is not essential, but merely convenient for the sake of
comparison with the AdS/CFT results.) In writing \eqnum{xGx}, we have
also divided by the nuclear transverse area $\pi R^2$, so that the
quantity shown there is the number of gluons per unit transverse area
having transverse momenta $k_\perp\lesssim Q < Q_s(x)$. Then, the
estimate in the r.h.s. of this equation can be understood as follows: the
gluons with longitudinal momentum fraction $x$ and low transverse momenta
$k_\perp < Q_s(x)$ are {\em saturated}, in the sense that there is a
number of gluons of $\order{1/\lambda}$ of each color per unit cell in
the phase--space, and this number does not grow with the energy anymore
(in contrast to what happens at larger momenta $k_\perp > Q_s(x)$).
Specifically, \eqnum{xGx} implies the following estimate for the gluon
occupation number at saturation in QCD at weak coupling and large $N_c$ :
 \beq
 n(x,k_\perp) \,\equiv\,\frac{1}{N_c^2 \pi R^2}\,\frac{\rmd \,xG(x,Q^2)}
 {\rmd Q^2}\bigg|_{Q=k_\perp}
 \,\sim\,\frac{1}{\lambda}\,\ln\frac{Q_s^2(x)}{k_\perp^2}\,,\eeq
which shows only a weak, logarithmic, dependence upon the energy, via the
corresponding dependence of the saturation momentum: $Q_s^2(x)\sim
1/x^\omega$ with $\omega\sim\order{\lambda}$.

We shall now argue that the gluon distribution in \eqnum{xGx} is the
natural quantity to compare with the structure function $F_2$ in
\eqnum{Flow} and that this comparison suggests a partonic interpretation
for the latter which is further supported by the sum rules
\eqref{sumrules}.

As already mentioned, the $\mcal{R}$--current directly couples to
fermionic and scalar fields of $\mcal{N}=4$ SYM, so the respective
structure function $F_2$ is most naturally related (at least at weak
coupling, where the standard OPE arguments apply) to the parton
distributions for these fields. However, the latter lie in the adjoint
representation of the color group; hence, for the purposes of the $N_c$
power counting, it is more appropriate to compare $F_2$ in $\mcal{N}=4$
SYM to the {\em gluon} distribution in QCD, and not to the respective
structure function, which rather describes the distribution of quarks in
the fundamental representation. Such a comparison, together with the
formal similarity between \eqnum{xGx} and the expression \eqref{Flow} for
$F_2$ (they both show the same parametric dependencies upon $N_c^2$ and
$Q^2$), immediately suggests that partons are saturated at strong
coupling too, but with occupation numbers of order one, rather than
$1/\lambda$. More precisely, for $k_\perp \ll Q_s(x)$, \eqnum{Flow}
implies
  \beq\label{nshw}
 n(x,k_\perp) \,\equiv\,\frac{\Lambda L }{N_c^2}\,\frac{\rmd F_2(x,Q^2)}
 {\rmd Q^2}\bigg|_{Q=k_\perp}
 \,\sim\,\ln\frac{Q_s^2(x)}{k_\perp^2}\,,\eeq
which suggests that, at strong coupling, the mechanism for parton
saturation has the same effect as at weak coupling \cite{CGCreviews}
(although its dynamical nature might be very different)
--- it limits the {\em rate} for parton emission with increasing rapidity
$Y\equiv \ln(1/x)$ : $\rmd n/\rmd Y\sim$~const. for $k_\perp\lesssim
Q_s(Y)$. At least at weak coupling, this limitation is associated with
the formation of a `color glass condensate'  \cite{CGCreviews} --- a
strong color mean field collectively created by all the small--$x$
partons, which blocks the radiation of new partons once the density is
high enough. It is likely that a similar picture holds at strong coupling
as well. But unlike at weak coupling, where this saturation requires
parametrically large occupation numbers $n\sim 1/\lambda \gg 1$, to
compensate for the weakness of the coupling, at strong coupling it occurs
already for occupation numbers of $\order{1}$, since the mutual repulsion
becomes strong as soon as the partons are allowed to interact with each
other.


In writing \eqnum{nshw} we have extracted the parton occupation number at
saturation out of the structure function $F_2$ in the same way as we
would do at weak coupling. To further justify this identification, let us
now show that it is also consistent with the sum rules \eqref{sumrules},
which in fact explain the additional factor $\Lambda L$ that we have
inserted in the definition of $n(x,k_\perp)$ in \eqnum{nshw}. We have
already mentioned below \eqnum{sumrules} that the following integral
 \beq
  \int_{0}^{1} \dif x\, F_2(x,Q^2) \,\sim\,x_s\, F_2(x_s,Q^2)
 \,\sim\,N_c^2\,\Lambda^2\,,
 \eeq
is proportional to the energy density per unit longitudinal distance in
the target rest frame. Using this, we would like to construct the energy
per unit transverse area in the shockwave\footnote{Of course, we already
know the corresponding result --- it reads $({\rmd E_{\rm
sw}}/{\rmd^2b})\sim \gamma N_c^2\,\Lambda^4 L$ ---, but here we are
rather interested in expressing this result in terms of $F_2$ and thus
deducing a partonic interpretation for the latter.}. This is obtained by
multiplying the quantity above by the longitudinal extent $L$ (to get the
total energy), then dividing it by the transverse area (\emph{i.e.}
multiplying it by a factor $\Lambda^2$), and finally multiplying it by
the Lorentz factor $\gamma$ (to make a boost to the target infinite
momentum frame). One thus gets
  \beq\frac{\rmd E_{\rm sw}}{\rmd^2b}
 \,\sim \,\gamma L\Lambda^2\,\big(x F_2(x,Q^2)\big)_{x=x_s}
 \,\sim \, k^+ \big(\Lambda L F_2(x,Q^2)\big)_{x=x_s}\,,
 \eeq
where we have also identified the longitudinal momentum
$k^+=xp^+=x\gamma\Lambda$ of the parton (scalar or adjoint fermion) which
absorbs the virtual photon in this frame. The last estimate above
suggests that, at least for $x\sim x_s$ where most of the energy is
located, the quantity $\Lambda L F_2(x,Q^2)$ has the meaning of the total
number of partons (with energy fraction $x$) per unit transverse area as
probed on the resolution scale $Q^2$:
 \beq
 \Lambda L F_2(x,Q^2)\,\sim\,\int\limits^{Q^2}\,\rmd^2k_\perp
 \ \frac{\rmd n}{\rmd^2k_\perp \rmd^2b_\perp}\,.\eeq
This is indeed the same as \eqnum{nshw} since, by definition,
$n(x,k_\perp)\equiv (1/N_c^2)\,[{\rmd n}/{\rmd^2k_\perp \rmd^2b_\perp}]$.

\section*{Acknowledgments}

We would like to thank Guillaume Beuf for early discussions which
motivated this work and also for pointing to us
Refs.~\cite{Podolsky:1997ik,Kaigorodov}. We are grateful to Al Mueller
for many inspiring discussions and comments on the manuscript. We
acknowledge useful discussions with Javier Albacete, Jorge Casalderrey--Solana and Tony Rebhan. The work of E.A. and E.I. is supported in part by
Agence Nationale de la Recherche via the programme ANR-06-BLAN-0285-01.
The research of  L.M. is supported under DOE Contract No. DE-
AC02-98CH10886. The work of D.N.T. is supported by the Austrian Science
Foundation FWF project no.~P19958 and in part by \"{O}AD project nos.~FR
08/2009 and ES 12/2009.

\appendix

\section{From black holes to shockwaves}
\label{app-bhsw}

Let us modify the temperature dependent piece of the AdS$_5$ black--brane
metric by an arbitrary function depending on the $x^3$ coordinate. That
is, using Fefferman--Graham coordinates, we consider the metric in
 \beq\label{bhmetric}
  \dif s^2 =
  \frac{1}{z^2} \left\{ -\frac{(1-a)^2}{1+a}\,
  \dif t^2 + (1+a)\left[\dif x_{\perp}^2 + (\dif x^3)^2\right] +
  \dif z^2 \right\} \quad \mathrm{with} \quad
   a \equiv c(x^3)\frac{z^4}{4 z_0^4}.
 \eeq
Here $c(x^3)$ is a dimensionless function, which reduces to unity for the
black--brane solution, and $z_0 = 1/\pi T$ with $T$ the Hawking
temperature. In general Eq.~\eqref{bhmetric} is not a solution to the
Einstein equations. Now we would like to make a boost and choose a frame
$\acute{x}^m$ which moves with velocity $-\upsilon$ with respect to the
frame $x^m$ and along their common axis in the 3-direction. In the new
frame we would also like to switch to light-cone coordinates and it is
straightforward to check that the combined transformation of coordinates
reads
 \beq\label{LorentzLCt}
  t \!\!&&= \frac{1}{\sqrt{2} \gamma (1+ \upsilon)}\, \acute{x}^{+} +
  \frac{\gamma (1+ \upsilon)}{\sqrt{2}}\, \acute{x}^{-},
  \\ \label{LorentzLCx}
  x^3 \!\!&&= \frac{1}{\sqrt{2} \gamma (1+ \upsilon)}\, \acute{x}^{+} -
  \frac{\gamma (1+ \upsilon)}{\sqrt{2}}\, \acute{x}^{-} .
 \eeq
Clearly the transverse and radial coordinates remain unaffected. The
metric tensor transforms according to
 \beq\label{gtrans}
  \acute{g}_{mn} = \frac{\del x^p}{\del \acute{x}^m}\,\frac{\del x^q}{\del \acute{x}^n}\,g_{pq},
 \eeq
so that the components which are transformed read
 \beq\label{g++}
 \acute{g}_{++} \!\!&&= \frac{1}{2 \gamma^2\, (1+\upsilon)^2}\,(g_{tt}+g_{33}) = \frac{2 a}{\gamma^2(1+\upsilon)^2 z^2 (1+a)},
 \\ \label{g--}
 \acute{g}_{--} \!\!&&= \frac{\gamma^2\, (1+\upsilon)^2}{2}\,(g_{tt}+g_{33}) = \frac{2 \gamma^2 (1+\upsilon)^2 a}{z^2 (1+a)},
 \\ \label{g+-}
 \acute{g}_{+-} \!\!&&= \acute{g}_{-+}= \frac{1}{2}\,(g_{tt} - g_{33}) = -\frac{(1+a^2)}{z^2 (1+a)}.
 \eeq
When the boost is ultra relativistic, that is for $\gamma \gg 1
\Leftrightarrow \upsilon \to 1$, we expect the metric to simplify. Indeed
we have
 \beq\label{gultra}
 \acute{g}_{++} \simeq 0, \quad
 \acute{g}_{--} \simeq \frac{1}{z^2\,}\frac{2 \gamma^2 \tilde{c}(\acute{x}^-)z^4}{z_0^4}\,\frac{1}{1+ \frac{\tilde{c}(\acute{x}^-) z^4}{4 z_0^4}} \quad \mathrm{and} \quad
 \acute{g}_{+-} = \acute{g}_{-+} = - \frac{1}{z^2}\,
 \frac{1+ \frac{[\tilde{c}(\acute{x}^-)]^2 z^8}{16 z_0^8}}
 {1+ \frac{\tilde{c}(\acute{x}^-) z^4}{4 z_0^4}},
 \eeq
where we have substituted $a$ with its explicit form and we have defined
$\tilde{c}(\acute{x}^-) = c(x^3)$. Notice that in this ultra relativistic
limit Eq.~\eqref{LorentzLCx} simplifies to $x^3 \simeq -\sqrt{2} \gamma
\acute{x}^-$. Furthermore, close to the boundary, i.e.~when $z \ll z_0$,
one can approximate $g_{ii} \simeq 1/z^2$ in Eq.~\eqref{bhmetric} and
also neglect the last factor in $\acute{g}_{--}$ and $\acute{g}_{+-}$ in
Eq.~\eqref{gultra}. However, notice that $\acute{g}_{--}$ does not vanish
since it is enhanced by the square of the large boost factor. Putting
everything together, and dropping the primes,  we can write the metric as
 \beq\label{swmetric}
  \dif s^2 \simeq \frac{1}{z^2} \left[
  \frac{2 \gamma^2 \tilde{c}(x^-) z^4}{z_0^4}\, (\dif x^-)^2 - 2 \dif x^+ \dif x^- + \dif x_{\perp}^2 + \dif z^2
  \right].
 \eeq
This is formally the same as the metric corresponding to a homogeneous in
the transverse space shockwave, cf. \eqnum{bhom}, which we recall is an
exact solution to the Einstein equations. Here, however, this metric
emerges only as a near--boundary approximation valid for $z\ll 1/T$. On
the other hand, the more general boosted--plasma metric in \eqnum{gultra}
keeps trace of the black hole horizon at $z=z_0$, and it is valid only
for $z\le z_0$. This makes it natural to supply the shockwave in
\eqnum{bhom} with a radial cutoff at finite $z \sim 1/T$ or in general at
$1/\Lambda$, where $\Lambda$ is an infrared cutoff in the dual gauge
theory.

Let us finally observe that when the function $c(x^3)$ has support only
in an interval of finite length $L$, in the boosted system it will be
transformed into a delta function. For instance, in the case of a slice
of a plasma (in the dual gauge theory) with $c(x^3) =1$ for $|x^3| \leq
L/2$ and zero otherwise, in the boosted frame and in light cone
coordinates we find
 \beq\label{tildec}
  \tilde{c}(x^-) \simeq \frac{L}{2 \sqrt{2} \gamma}\, \delta(x^-).
 \eeq

\section{The bulk--to--bulk propagator}
\label{app-G}

In this Appendix we outline the construction of the Green's functions for
the equations of motion. For definiteness, let us consider the dilaton
case. Taking the Fourier transform of \eqref{eqG} w.r.t.~$x^-$ and
$x_\perp$ we have
 \beq
 \Big(\partial_z^2 - \frac{3}{z}\,\partial_z  - K^2\Big)
 G(z,\acute{z};K^2) = z^3\delta(z-\acute{z}),
 \eeq
where we have defined the variable $K^2 \equiv k_\perp^2 - 2q^-k^+$. The
solution to the eigenvalue problem for the operator $\del_z^2 -
(3/z)\,\del_z$ is given by
 \beq
 \Big(\partial_z^2 - \frac{3}{z}\,\partial_z\Big) z^2 \, \rmJ_2(\omega z) =
 -\omega^2 z^2 \, \rmJ_2(\omega z),
 \eeq
and the eigenfunctions satisfy the completeness relation
 \beq\label{complet}
 \int_0^\infty \rmd\omega \,
 \omega\, z^2 \rmJ_2(\omega z)\, \acute{z}^2 \rmJ_2(\omega\acute{z}) =
 z^3 \delta(z-\acute{z}).
 \eeq
Therefore we can write the Green's function as
 \beq\label{green}
 G(z,\acute{z};K^2) = - \int_0^\infty \rmd\omega \, \omega\,
 \frac{1}{\omega^2 + K^2}\, z^2 \rmJ_2(\omega z) \, \acute{z}^2
 \rmJ_2(\omega\acute{z}).
 \eeq
Now we would like to integrate over $\omega$ and we need to distinguish
two possibilities according to the sign of $K^2$. In the spacelike case
$K^2>0$ we find
 \beq\label{Gspace}
 G(z,\acute{z};K^2>0) =
 -z^2 \acute{z}^2 [ \rmK_2(K\acute{z})\rmI_2(K z) \Theta(\acute{z}-z)
 + \rmK_2(K z) \rmI_2(K\acute{z}) \Theta(z-\acute{z})].
 \eeq
In the timelike case $K^2<0$ we need a prescription to avoid hitting the
pole which is located on the positive $\omega$-axis. Letting $K^2 \to K^2
- \rmi \epsilon$ with $\epsilon >0 $, we find (in agreement with the
analytic continuation of \eqref{Gspace})
 \beq\label{Gtime}
 G(z,\acute{z};K^2<0) = -\frac{\rmi \pi z^2 \acute{z}^2}{2}
 [\rmH_2^{(1)}(|K| \acute{z}) \rmJ_2(|K| z) \Theta(\acute{z}-z) +
  \rmH_2^{(1)}(|K| z) \rmJ_2(|K| \acute{z}) \Theta(z-\acute{z}) ].
\eeq
Following the same procedure we can construct the Green's functions that
we encountered for the $\mcal{R}$--current in Sect.~\ref{sect-R}. The
transverse and longitudinal Green's functions in equations \eqref{GT1}
and \eqref{GL1} are given by
 \beq\label{GT}
 &&G_{\rm T}(z,\acute{z}; K^2>0) = - z \acute{z} [
 \rmK_1(K\acute{z}) \rmI_1(K z)\Theta(\acute{z} -z) +
 \rmK_1(K z) \rmI_1(K \acute{z})\Theta(z- \acute{z})],
 \\ \label{GL}
 &&G_{\rm L}(z,\acute{z}; K^2>0) = - z \acute{z} [
 \rmK_0(K\acute{z}) \rmI_0(K z)\Theta(\acute{z} -z) +
 \rmK_0(K z) \rmI_0(K \acute{z})\Theta(z- \acute{z})],
 \eeq
respectively. The corresponding functions for $K^2<0$ are obtained via
analytic continuation. Now the  Green's functions in coordinate space
will be given by
 \beq
 G(z,\acute{z}; x^- - y^-, x_{\perp} - y_{\perp}) =
 \int \frac{\dif k^+ \dif^2 k_{\perp}}{(2\pi)^3} \,
 \rme^{- \rmi k^+(x^- - y^-) + \rmi k_{\perp} \cdot
 (x_{\perp} - y_{\perp})} G(z,\acute{z}; K^2).
 \eeq
and the integration over $k^+$ can be easily performed using the
representation in Eq.~\eqref{green} with our prescription $K^2 \to K^2 -
\rmi \epsilon$. The pole is at $(\omega^2 + k_{\perp}^2)/2 q^- - \rmi
\epsilon$ and is enclosed by the integration contour only for $x^->y^-$.
Thus the $k^+$-integration leads to the factor
 \beq
 -\int \frac{\dif k^+}{2 \pi}\,\frac{\rme^{- \rmi k^+(x^- - y^-)}}{\omega^2 + k_\perp^2 - 2 q^- k^+ - \rmi \epsilon} = - \frac{\rmi \Theta(x^- - y^-)}{2 q^-}\, \exp\left[-\frac{\rmi (x^-- y^-)
  (\omega^2 + k_{\perp}^2)}{2 q^-}\right],
 \eeq
which in turn yields the expression in \eqnum{GR}.

The bulk--to--bulk propagator has been already constructed in the past
(see for example \cite{Danielsson:1998wt,D'Hoker:2002aw,Lin:2009pn}), but
mostly using Euclidean signature, or using standard Minkowski coordinates
instead of light cone coordinates. Therefore, let us now show how the
methods used in this paper lead to these known results. We will aim to
write the propagator as a function of the chordal distance introduced
earlier in Sect.~\ref{sect-SW}. This property arises from the fact that
the differential operator can be rewritten as the Laplacian of the
hyperbolic space $\mathbb{H}_{d+1}$ with coordinates $(z,x_1, \dots,
x_d)$. We shall be interested in the case $d=4$. Defining the
$4$--dimensional vector $r = x - y$ and performing a Wick's rotation in
Eq.~\eqref{green}, the Euclidean Green's function in coordinate space is
obtained as (below, $k^2=\sum_{i=1}^4k_i^2$ is strictly positive)
 \beq
 G(z,\acute{z};r) = -z^2 \acute{z}^2
 \int \frac{\rmd^4 k}{(2\pi)^4}\, \rme^{\rmi k\cdot r}
 \int_0^\infty \rmd\omega \, \omega\, \frac{1}{\omega^2+k^2} \,
 \rmJ_2(\omega z)\rmJ_2(\omega \acute{z}).
 \eeq
First we perform the integration over $k$. In $d$--dimensions we have
 \beq\label{FTprop}
 \int \frac{\rmd^d k}{(2\pi)^d}\,
 \frac{\rme^{\rmi k\cdot r}}{\omega^2+k^2} = \frac{1}{(2 \pi)^{d/2}}\,
 \left(\frac{\omega}{r}\right)^{d/2-1}\,
 \rmK_{d/2-1}(\omega r).
 \eeq
Using the above for $d=4$ and the identity $\rmK_1(\omega r) =
-(1/\omega) \del_r \rmK_0(\omega r)$, we can write the coordinate space
propagator as
\beq\label{Gcoord}
 G(z,\acute{z};r) =
 \frac{z^2 \acute{z}^2}{4 \pi^2 r}\, \frac{\del}{\del r}
 \int_0^\infty \rmd \omega \, \omega
 \rmJ_2(\omega z) \rmJ_2(\omega\acute{z}) \rmK_0(\omega r) =
 \frac{z^2 \acute{z}^2}{4 \pi^2 r}\, \frac{\del}{\del r}\,
 \frac{(r_2 - r_1)^2}{r_1 r_2 (r_2+r_1)^2}
 \eeq
where we have found convenient to define
 \beq
 r_1 = \sqrt{r^2 + (z-\acute{z})^2}
 \quad \mathrm{and} \quad
 r_2 = \sqrt{r^2 + (z+\acute{z})^2}.
 \eeq
Now we introduce a chordal coordinate $\xi$, which is closely related to
the chordal distance $q$ defined in Sect.~\ref{sect-SW}, according to
 \beq\label{chordalxi}
 \xi = \frac{2 z \acute{z}}{r^2 + z^2 + \acute{z}^2} = \frac{1}{1+2 q}.
 \eeq
It is straightforward to make a change of variable from $r$ to $\xi$ and
performing the required algebraic manipulations we finally arrive at
 \beq
 G(\xi) = -\frac{1}{8\pi^2}\,
 \frac{\xi^4 (1+ 2 \sqrt{1-\xi^2})}
 {(1-\xi^2)^{3/2} (1+\sqrt{1-\xi^2})^2} =
 -\frac{3\xi^4}{32 \pi^2}\,
 _2F_1(2,5/2,3;\xi^2),
 \eeq
where $_2 F_1$ is the hypergeometric function. As announced, this
expression depends on the variables $z$, $\acute{z}$ and $r$ only through
a single chordal coordinate. Up to a minus sign, which is due to
different conventions, this is equal to the bulk--to--bulk propagator
given in \cite{D'Hoker:2002aw}\footnote{One needs to set $\Delta = d = 4$
in Eq.~(6.12) of that reference in order to obtain the 5-dimensional
massless case.}.

Furthermore, let us choose the `transverse' space to be 2--dimensional.
By making use of Eq.~\eqref{FTprop} for $d=2$, we can write the
propagator as
 \beq
 G_2(z,\acute{z};r) =
 -\frac{z^2 \acute{z}^2}{2 \pi r}
 \int_0^\infty \rmd \omega \, \omega
 \rmJ_2(\omega z) \rmJ_2(\omega\acute{z}) \rmK_0(\omega r) =
 \frac{z^2 \acute{z}^2}{4 \pi^2 r}\,
 \frac{(r_2 - r_1)^2}{r_1 r_2 (r_2+r_1)^2}.
 \eeq
We can easily  express the above in terms of the chordal distance $q$ to
find
 \beq\label{G2}
 G_2(z,\acute{z};q) = -\frac{z \acute{z}}{128\pi q^3}\,
 _2 F_1(3,5/2,5;-1/q).
 \eeq
Multiplying the above by $8 \pi^2 E \delta(x^-)/N_c^2$, as dictated from
Eqs.~\eqref{Jbulk} and \eqref{bequation}, we are lead to
Eq.~\eqref{bgubser}. Finally, yet another definition is to use $\cosh u =
1/\xi$, which means that we can write
 \beq
 G_2(z,\acute{z};u) = - \frac{z \acute{z}}{4\pi}\,
 \frac{\rme^{-2 u}}{\sinh u},
 \eeq
which, up to an overall factor independent of $u$, is the form used in
\cite{Lin:2009pn}\footnote{In Eq.~(25) in that reference, where the sign
in the exponent should be negative.}.

\section{Prefactor of the structure function}
\label{app-pre}

In this short Appendix we give an alternative calculation of the
structure function at saturation in the limit $Q^2 \ll Q_s^2(x,0)$, in
the case of probing with a dilaton, and in order to determine the correct
prefactor. We can write the most dominant contribution to the structure
function as
 \beq
 F(x,Q^2) = \frac{N_c^2}{32 \pi^2}\,\frac{Q^6 \Lambda}{x L}
 \int_{0}^{R} \dif^2 \bp \int_{0}^{\infty} \dif z \,
 z\, \rmK_2^2(Q z)
 \left\{1 - \cos\left[Q^2 Q_s^2(x,\bp)\,z^4\right] \right\},
 \eeq
where the radius $R$ is defined by $Q_s^2(x,R) = Q^2$. We have already
seen in Sect.~\ref{sect-sf} that the dominant contribution in this
dilaton case comes from the region $z \ll 1/Q$, thus it is natural to
approximate the Bessel function by its lowest order term, that is
$\rmK_2(Q z) \simeq 2/(Q z)^2$. Then the $z$-integration gives
$\sqrt{2\pi}\, Q_s(x,\bp)/Q^3$ and the structure function becomes
 \beq\label{sfexact}
 F(x,Q^2) = \frac{N_c^2}{16 \sqrt{2\pi}}\,\frac{Q^4 }{x \Lambda L}\,
 \frac{Q_s(x,0)}{Q}\,
 \underbrace{\frac{\Lambda^2}{\pi}\int_{0}^{\infty}
 \dif^2 \bp \sqrt{\frac{f(\bp \Lambda)}{f(0)}}}_{\kappa}.
 \eeq
Notice that we have replaced the radius $R$ in the upper limit of the
impact parameter integration by $\infty$. This happens because the
presence of the profile function in the numerator of the integrand
effectively restricts the integration at values $\bp \lesssim 1/\Lambda$,
while at the same time $R \gg 1/\Lambda$. {\it A posteriori} this also
justifies the small-$z$ approximation since for all values of $\bp$ that
contribute, one has $Q^2 \ll Q_s^2(x,\bp)$. Now, $\kappa$ in
Eq.~\eqref{sfexact} is a pure number which depends on the profile of the
shockwave. For instance, $\kappa=8$ for the exponential type-I shockwave
of the form $f(\bp \Lambda) = (1/2 \pi) \exp(-\bp\Lambda)$, while
$\kappa=2$ for the type-II one.

\section{Explicit check of sum rules}
\label{app-sr}

Here we will explicitly check the validity of the sum rule for the
structure function $F_1$ as stated in the first expression in
Eq.~\eqref{sumrules}. One can follow a similar procedure for other sum
rules. Using Eqs.~\eqref{FT} and \eqref{TmQs} which provide us with $F_1$
and the scattering amplitude $\mcal{T}$ respectively we have
 \beq
 F_1(x,Q^4) = \frac{N_c^2}{32 \pi^3}\,\frac{Q^4}{x \Lambda L}
 \int_{0}^{\infty} \dif z\, z\, \rmK_1^2(Q z)
 \left[1 - \cos\left(Q^2 Q_s^2\,z^4\right) \right].
 \eeq
We wish to integrate the above from $x=0$ to $x=1$ which is the physical
region. Making a change of variables from $z$ to $\zeta = Q z$ and from
$x$ to $t=Q_s^2(x)/Q^2$, we can write
 \beq
 \int_0^1 \dif x\, x F_1(x,Q^2) = \frac{N_c^2 \Lambda^2}{64 \pi}
 \int_{0}^{\infty} \dif \zeta\, \zeta\, \rmK_1^2(\zeta)
 \int_{t_1}^{\infty} \dif t\,\frac{1 - \cos t \zeta^4}{t^2},
 \eeq
with $t_1 = Q_s^2(1)/Q^2 = \pi^2 \Lambda^3 L/2 Q^2$. It is
straightforward to see that the sum rule will be exactly recovered only
when we set $t_1 \to 0$ as the lower limit in the $t$-integration. For
small but non-vanishing $t_1$, we can separate an integration from $0$ to
$\infty$ which gives the sum rule, minus an integration from $0$ to $t_1
\ll 1$ for which we can do a Taylor expansion. We find
 \beq
 \int_0^1 \dif x\, x F_1(x,Q^2) = \frac{N_c^2 \Lambda^2}{80} -
 \frac{8 N_c^2 \Lambda^2}{7 \pi}\,\frac{Q_s^2(1)}{Q^2} + \cdots.
 \eeq


\providecommand{\href}[2]{#2}\begingroup\raggedright\endgroup

\end{document}